%% file: proof-of-vax.tex
\documentclass[USenglish,oneside]{article}

\usepackage{arxiv}

\usepackage[utf8]{inputenc}

\usepackage{amsmath}
\usepackage{paralist}
\usepackage{algorithm}
\usepackage{balance}
\usepackage{xspace}
\usepackage{adjustbox}
\usepackage{multirow}
\usepackage{footnote}
\usepackage{url}
\usepackage{booktabs}
\usepackage{wasysym}
\usepackage[group-separator={,},binary-units,per-mode=symbol,group-minimum-digits=4]{siunitx}
\usepackage{xcolor}
\usepackage{booktabs}

\usepackage{caption}
\usepackage{subcaption}

\newcommand{\cf}{cf.\@\xspace}
\newcommand{\ie}{i.\@\,e.\@\xspace}
\newcommand{\eg}{e.\@\,g.\@\xspace}
\newcommand{\wrt}{w.\@\,r.\@\,t.\@\xspace}

\newcommand{\etal}{et~al.\@\xspace}

\begin{document}
  
  \author{
  Marvin Kowalewski\\
  Ruhr University Bochum, Germany\\
  \texttt{marvin.kowalewski@rub.de}
  \and
  Franziska Herbert\\
  Ruhr University Bochum, Germany\\
  \texttt{franziska.herbert@rub.de}
  \and \\
  Theodor Schnitzler\\
  Ruhr University Bochum, Germany\\
  \texttt{theodor.schnitzler@rub.de}
   \and \\
  Markus Dürmuth\\
  Ruhr University Bochum, Germany\\
  \texttt{markus.duermuth@rub.de}
}

  \title{\huge Proof-of-Vax: Studying User Preferences and Perception of Covid Vaccination Certificates}
  \rhead{\scshape Proof-of-Vax: Studying User Preferences and Perception of Covid Vaccination Certificates}

\maketitle

\begin{abstract}
{
Digital tools play an important role in fighting the current global COVID-19 pandemic.
We conducted a representative online study in Germany on a sample of 599 participants to evaluate the user perception of vaccination certificates.
We investigated five different variants of vaccination certificates, based on deployed and planned designs in a between-group design, including paper-based and app-based variants.
Our main results show that the willingness to use and adopt vaccination certificates is generally high. Overall, paper-based vaccination certificates were favored over app-based solutions. 
The willingness to use digital apps decreased significantly by a higher disposition to privacy, and increased by higher worries about the pandemic and acceptance of the coronavirus vaccination.
Vaccination certificates resemble an interesting use case for studying privacy perceptions for health related data. We hope that our work will be able to educate the currently ongoing design of vaccination certificates, will give us deeper insights into privacy of health-related data and apps, and prepare us for future potential applications of vaccination certificates and health apps in general.
}
\end{abstract}

\input{sections/1-introduction}
\input{sections/2-background}

\input{sections/3-method}

\input{sections/4-results}

\input{sections/5-discussion}

\input{sections/6-limitations}
\input{sections/7-conclusion}

\bibliographystyle{IEEEtranS}
\bibliography{proof-of-vax}

\appendix

\input{sections/questionnaire}

\end{document}

%% file: sections/1-introduction.tex
\section{Introduction}

The global pandemic caused by the Severe Acute Respiratory Syndrome Coronavirus~2 (SARS-CoV-2) has hit the world in early 2020. 
This led to worldwide restrictions in social life, freedom of travel, and caused severe damage to the global economy. 
In the course of this pandemic, different measures were used, including strategies to contain the coronavirus with the help of digital tools.
Several countries have rolled out mobile apps for different purposes such as contact tracing~\cite{horowitz_europe_2020} or quarantine monitoring~\cite{davidson_healthcode_2020}.

The development of various vaccines against COVID-19 along with the continuous vaccination progresses in different countries has sparked discussions around how vaccination certificates should be realized~~\cite{mathieu21:global-database-covid, mathieu21:vax-data}.
Such certificates can facilitate the withdrawal of lockdown restrictions particularly for vaccinated people~\cite{germandebateprivileges_2021}.
There are currently different digital and non-digital approaches used around the world.
In the UK and the US, paper-based vaccination cards are used to prove the vaccination status~\cite{cdc_ukproofofvax_2020, us21:vaccination-record-us}.
Israel, a country with a comparably fast vaccination progress, was one of the first countries to introduce a so-called \emph{Green Pass}, both as a paper-based certificate and a digital app, granting specific privileges to vaccinated citizens~\cite{israel21:digital-green-pass}.
The European Union is planning to release the \emph{Digital Green Certificate} as a universal solution that is recognized throughout Europe in this summer~\cite{eu21:digital-green-certificate}.

In this work, we present the results of an online study conducted in Germany in March 2021 with \num{599} participants selected representatively for the German population \wrt age, gender, and education.
Our study was designed to capture users' preferences and perception of different types of COVID-19 vaccination certificates.
We study different types of vaccination certificates, comparing
\begin{inparaenum}[(i)]
\item digital apps to paper-based variants, and 
\item certificates specifically tailored to COVID-19 vaccinations to more general solutions that vaccination information can be integrated into.
\end{inparaenum}
This leads us to a total of five certificate variants, two paper-based and three digital variants, that we compare in a between-subject design with approx.\ 120 participants per condition.
We also evaluate the extent to which participants in our study would use the certificates for specific purposes.
More specifically, we consider within-subject use cases in which certificates are only used for documentation, compared to situations in which the certificate is required to attend specific activities.
For the evaluation of our study results, we follow a mixed-methods approach comprising quantitative measurements and qualitative analyses of participants' feedback.
In particular, we examine participants' willingness to use such vaccination certificates, the perceived usefulness or utility, and the perceived effort required for using the certificate.
Furthermore, we explore potential privacy concerns, particularly related to sensitive personal health data.

As the main results of our study show, participants are in general highly willing to use all variants of vaccination certificates and all vaccination certificate variants are rated as useful and easy to use with low effort.
For the documentation-only purpose, participants are significantly more willing to use paper-based than app-based variants.
We observe significant negative correlations between participants' disposition to privacy and their willingness to use vaccination certificates.
Aspects that significantly increase participants' willingness to use a vaccination certificates include
\begin{inparaenum}[(i)]
\item whether participants already use the official German contact tracing Corona-Warn-App, 
\item worries about getting infected with coronavirus, and
\item their attitude towards vaccinations in general.
\end{inparaenum}
On the downside, participants who oppose vaccination obligations are significantly less willing to use vaccination certificates.

The results of this study allow us to better understand users' rationale for accepting or rejecting the use of different forms of vaccination certificates.  
Even beyond the current pandemic, our results can serve as vital information for comparable situations in the future, where similarly vaccinations may play a role.  
In addition, our results allow for interesting insights into general privacy considerations of users, for the increasingly important area of digital health information apps~\cite{europe21:importanceofdigitalhealthdata}.

In summary, our research makes the following key contributions:
\begin{enumerate}
    \item We complement existing knowledge about acceptance of technology use in the current pandemic situation affecting a large number of people all over the world by expanding the view to vaccination certificates.
    \item Through quantitative and qualitative evaluations we show that privacy is an important factor for the adoption of digital tools processing highly sensitive personal data such as health information. 
    \item Our results provide insights that can support the upcoming design, development, and roll-out of vaccination certificates.
\end{enumerate}

%% file: sections/2-background.tex
\section{Related Work}
We describe work related to our study focusing on research that explores factors driving the acceptance of technology from general mobile apps and tools, to health apps, to solutions specifically tailored to help contain the current pandemic. 

\paragraph*{Technology Acceptance and Privacy in Digital Tools}
Literature on technology acceptance identified factors like perceived usefulness, perceived ease of use, social influence and demographic factors as gender and age to impact the acceptance and use of technology~\cite{davis_informationtechnology_1989, venkatesh_acceptance_2003}. 

Various studies have found that privacy plays an important role in decision-making about digital tools and interacting with online technology in a broad range of applications~\cite{emami-naeini21:which-privacy-security,linsner21:role-privacy-digitalization,schnitzler20:managing-longitudinal-privacy}.
Ray~\etal~\cite{ray20:warn-them-just} study privacy concerns among adults of two different age groups, finding that older adults are more concerned about global privacy threats.
Besides privacy being a factor influencing the use of a specific tool, there are also reasons for users to adopt privacy-enhancing technologies such as VPN or Tor to increase their privacy~\cite{harborth19:explaining-technology-use, namara19:emotional-practical-considerations}.
However, Story~\etal~\cite{story21:awareness-adoption-misconceptions} found severe misconceptions in the perception of protection provided by, \eg, VPN software.

\paragraph*{Factors Influencing the Use of Mobile Health Apps}
Research on sharing health information online or via apps has shown that users are more comfortable sharing health related data (\eg sleep, physical activity) with their doctors than with family members or electronic health records~\cite{nicholas_mentalhealth_2019}.
Bol~\etal~\cite{bol19:differences-mobile-health} examined which factors influence the use of health apps and found privacy to not be a significant factor. 
Differences in mobile health apps use were instead influenced by age, education level and e-health literacy.
In contrast, other works found users to have privacy concerns when using mobile health apps~\cite{gu_privacy_2017,wottrich_privacy_2018,zhou_barriers_2019}, or that they feared their data being misused by a third party and expressed desire for some level of control over their data~\cite{baig_dnatesting_2020}.
Concerning older participants, Rasche~\etal~\cite{rasche18:prevalence-health-app} found one sixth of older adults in Germany to use health apps, mostly exercise-related. 
Barriers for the use of such apps included lack of trust, privacy concerns, fear of misdiagnosis, and poor usability.
Related to usability issues, Peng~\etal~\cite{peng_healthapps_2016} found lack of app literacy, and lack of time and effort to be barriers for continued use of health apps.

\paragraph*{The Role Of Mobile Apps in the Current Pandemic}
The majority of research studying the role of mobile apps to contain the pandemic focus on solutions for digital contact tracing~\cite{zhang_covid_privacy_2020, altmann_acceptability_2020, simko_contact_tracing_2020, li_covid_apps_2020, trang_oneapp_2020, kaptchuk_covid19apps_2020,lu21:comparing-perspectives}.
Regarding other types of apps, Tsai~\etal~\cite{tsai21:exploring-promoting-diagnostic} explore users' needs for explanations and transparency in symptom checking apps.
Zhang~\etal~\cite{zhang21:mapping-landscape} developed a framework to systematize and analyze applications for public information about the pandemic, particularly focusing on different forms of visualizations.
Utz~\etal~\cite{utz21:apps-against-spread} expand the view to different types of apps used for various purposes, finding that privacy aspects such as data receivers but also the degree of anonymity significantly influence the acceptance of corona-related apps.

\section{Vaccination Certificates}\label{sec:vaxcert} 
We introduce different types of vaccination certificates as they are currently discussed or have already been rolled-out in different countries. 
We do so to motivate the selection of certificate realizations we use in our study.
Furthermore, we describe the state of the pandemic and vaccination progress in Germany at the time we conducted our study, in order to put the results into the context of the overall situation.

\subsection{State of the Art}
Vaccines are considered as one of the most effective measures to contain the pandemic noticeably. To prove that vaccination has taken place, different variants of proof of vaccination against the coronavirus, further called vaccination certificates, are used in several countries. To better understand the examined vaccination certificates used in this study, various forms of existing or planned vaccination certificates and their support in the fight against the current pandemic are presented below in more detail.

In the UK, a small wallet-size vaccination card published by the British health officials is used to prove the vaccination status. These paper-based vaccination cards include the citizens' name, the name of the vaccine, the unique batch number and the date of both vaccines given~\cite{cdc_ukproofofvax_2020}.
In the US, a similar COVID-19 vaccination record card is used and is the only proof of vaccination most Americans have after getting their COVID-19 shots. As of mid May 2021, the Centers for Disease Control and Prevention (CDC) stated that fully vaccinated citizens can resume activities without wearing a mask or physically distancing in certain situations. However, showing a negative test result or documentation of recovery from COVID-19 is still required before boarding an international flight to the US~\cite{us21:vaccination-record-us, us21:fully-vaccinated-privileges}.

Israel, a country with a very advanced vaccination campaign compared to other countries, was one of the first countries to introduce the so-called \emph{Green Pass} which is also published by official authorities~\cite{israel21:digital-green-pass}.
The issued vaccination certificate can be used within an app but can also be printed out in paper form. Personal information about the vaccination status can be read by means of a simple QR-code. In addition to the QR-code, the Green Pass shows the number of the personal ID card, the date of birth, the dates of vaccinations, and the vaccine's name~\cite{israel21:digital-green-pass}.
The green passport is being used to gradually re-open the country. Vaccinated citizens are allowed to visit gyms, hotels, theaters, or sport events. Israeli citizens are also allowed to enter the country without going through quarantine after a stay abroad~\cite{rasche21:green-pass-privileges}.

Not only governments but also companies started to develop vaccination certificates. 
The  \emph{International Air Transport Association} (IATA), representing 82\% of total air traffic, published the mobile app \emph{IATA Travel Pass}~\cite{iata21:digital-travel-pass}.
This mobile vaccination certificate is intended primarily to verify the vaccination status so that a passenger meets the requirements for traveling. It helps travelers to store and manage their verified certifications for COVID-19 vaccines~\cite{iata21:digital-travel-pass}.

In Germany, there are two possibilities to record the successful COVID-19 vaccination. The vaccination can be entered in the personal \emph{International Certificates of Vaccination} document standardized by the \emph{World Health Organization} (WHO). This yellow paper-based document for various vaccinations is internationally recognized in order to certify that vaccinations mandatory for travel and entering a specific country have been taken. The WHO Certificate is also required for verification purposes of certain vaccinations within countries. Unless this vaccination card is brought to the vaccination, a corona-specific paper-based vaccination certificate will be issued~\cite{vaccinepassport_2021, internationalcertificatesofvax_2021}.

In particular in Germany, a wide-spread and intensive debate related to privileges under the use of a vaccination certificate was and is still ongoing~\cite{germandebateprivileges_2021, germandebateprivilegesunfair_2021, rasche21:green-pass-privileges, us21:fully-vaccinated-privileges}.
This discussion gave rise to the idea of an Europe-wide vaccination certificate against COVID-19, the \textit{Digital Green Certificate}. The EU aims to introduce this kind of digital certificate valid throughout Europe by summer 2021~\cite{europeandigitalvaxcertificate_2021, germandebateprivilegesunfair_2021}.
The app-based certificate is primarily intended as a digital document but will also be available on paper associated with a personalized QR-code~\cite{europeandigitalvaxcertificate_2021}.
In order to speed up the tendering process, the \emph{Federal Ministry of Health} in Germany develops a digital certificate on its own, the \emph{CovPass App}. Just like the popular German contact tracing app, the \emph{Corona-Warn-App}, which is also usable with various other European contact tracing apps, the CovPass App is supposed to be compatible with the EU digital vaccination certificate. In addition, the digital vaccination certificate will be integrated into the existing Corona-Warn-App. Both apps will contain the same integrated QR-code protected by a digital signature. To verify the vaccination status, an additional verification app is developed~\cite{rki_downloadzahlen_2020, sap_coronawarnapp_2020, covpass_2021}.

\subsection{Current Situation in Germany}
% ==============================
% moved here from method section
% =============================
At the time this study was conducted, there was a second lockdown (\eg, shops closed, curfew at night, travel restrictions) due to significantly increasing infection numbers~\cite{rki_covid19dashboard_2020, germanysecondlockdownpoorly_2020}. 
There were approx. 150 infected per \num{100000} people on a weekly basis (\emph{7-days incidence rate}). Even though vaccination in Germany had started in December 2020, there was and still is limited access to vaccine both in Germany and worldwide~\cite{gerlatestcovidupdates_2020}.
As a result, only a small proportion of the population had been offered a COVID-19 vaccine. 
During the survey in Germany, from March 23 to 27 2021, only 4.5\% of the German population had been fully vaccinated~\cite{mathieu21:vax-data, germanyvaccinemonitoring_2021}.
Even before the study was conducted, public debates about societal implications of privileges for vaccinated citizens sparked in Germany.
Most discussed were withdrawals of coronavirus-related restrictions, especially for vaccinated people~\cite{germandebateprivileges_2021, germandebateprivilegesunfair_2021}. 
In addition to that, the form of vaccination certificates as well as proposals for an EU-wide uniform digital vaccination certificate were present in the media~\cite{europeandigitalvaxcertificate_2021}. 
Debates regarding especially digital solutions were dominated by privacy and security concerns of government-backed health applications~\cite{germandebateprivilegesunfair_2021, cccsocialimplications_2021}.

%% file: sections/3-method.tex
\section{Methodology}
\label{sec:methodology}

To investigate the general willingness to use of vaccination certificates, paper- and app-bases variants which contain sensitive health data, we conducted an online survey. We examined five different variants of vaccination certificates in a between-subject design. Additionally, we have retrieved three different use cases (within-subject design) for vaccination certificates per condition. A specific vaccination certificate combined with a use case is called a \emph{scenario}.

In the following, we describe our five different variants of vaccination certificates (between-subject), the associated three use cases (within-subject), and the survey questionnaire in detail.

\begin{figure*}[ht]
\centering
     \begin{subfigure}[b]{0.245\textwidth}
         \centering
         \begin{subfigure}[t]{\textwidth}
             \centering
             \includegraphics[width=\textwidth]{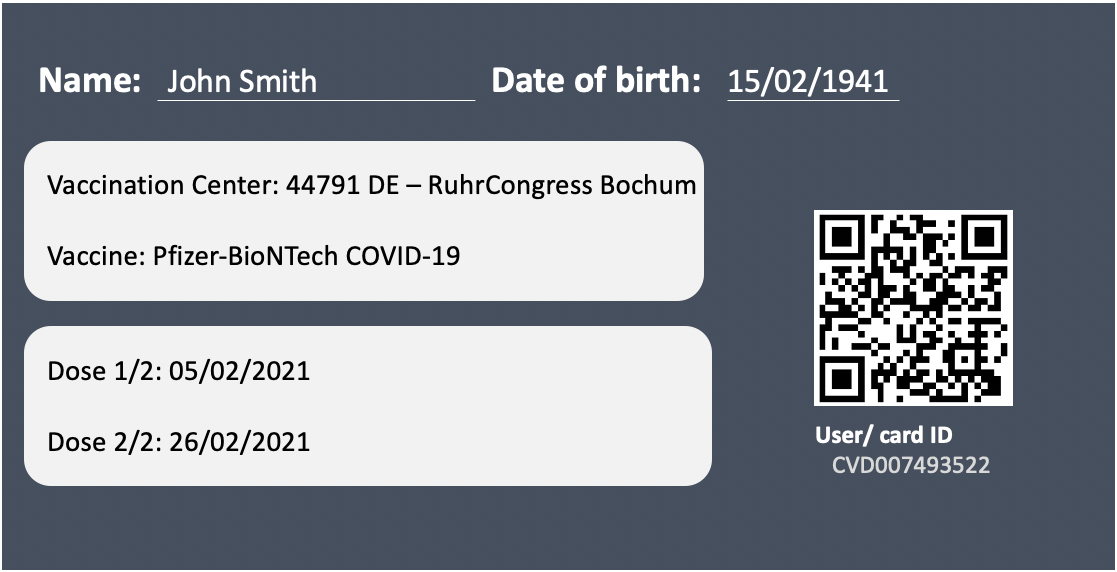}
             \caption{Covid Certificate (C1)}
             \label{fig:mockup-c1}
         \end{subfigure}
         \begin{subfigure}[b]{\textwidth}
             \centering
             \includegraphics[width=\textwidth]{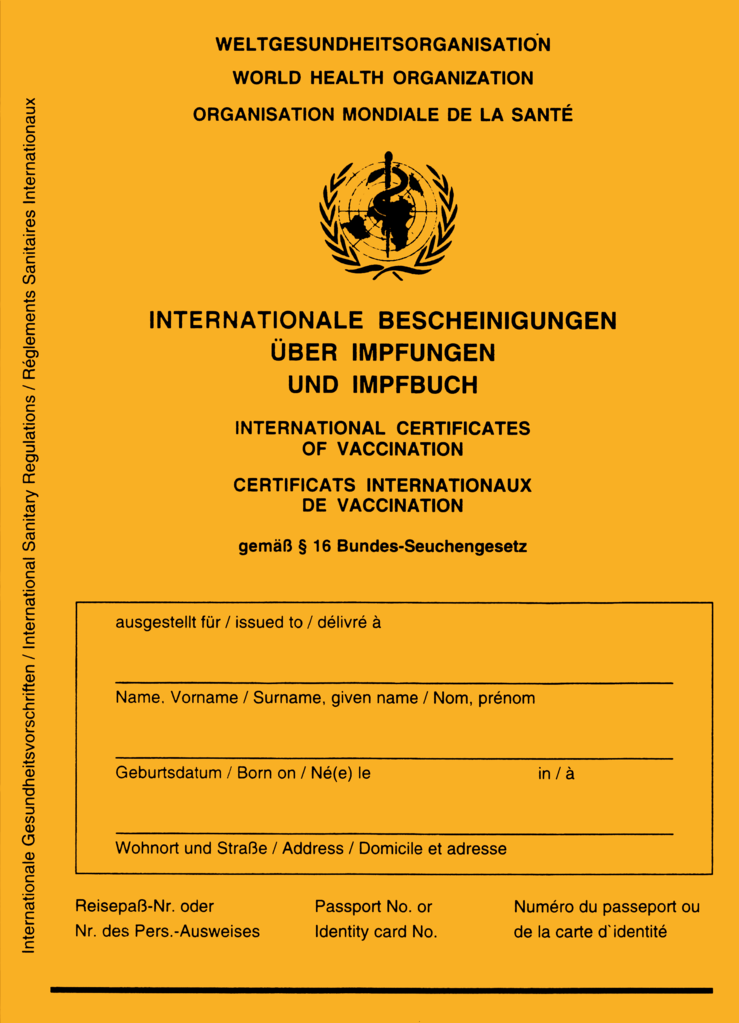}
             \caption{Intl. Certificate (C2)}
             \label{fig:mockup-c2}
        \end{subfigure}
     \end{subfigure}
     \begin{subfigure}[b]{0.245\textwidth}
         \centering
         \includegraphics[width=\textwidth]{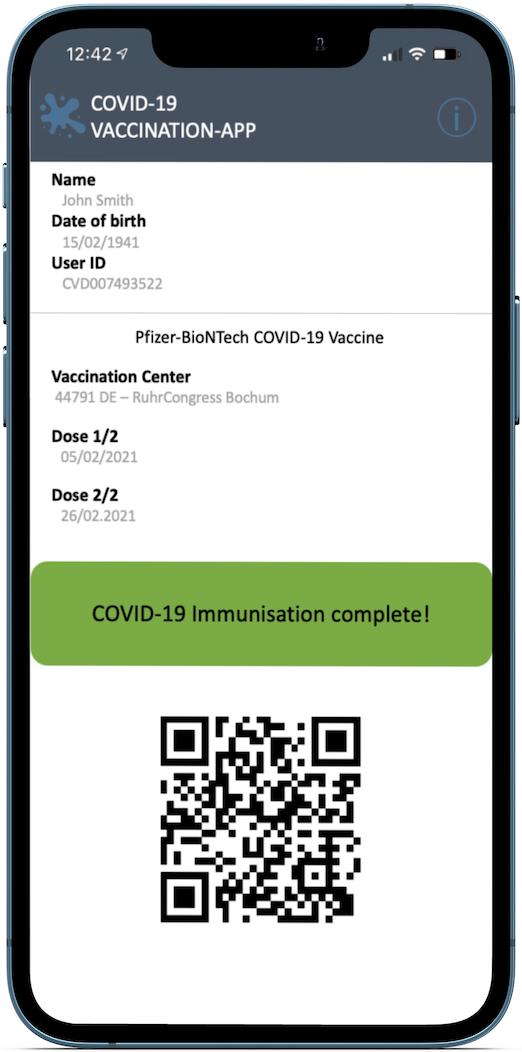}
         \caption{Covid Cert. App (C3)}
         \label{fig:mockup-c3}
     \end{subfigure}
     \begin{subfigure}[b]{0.245\textwidth}
         \centering
         \includegraphics[width=\textwidth]{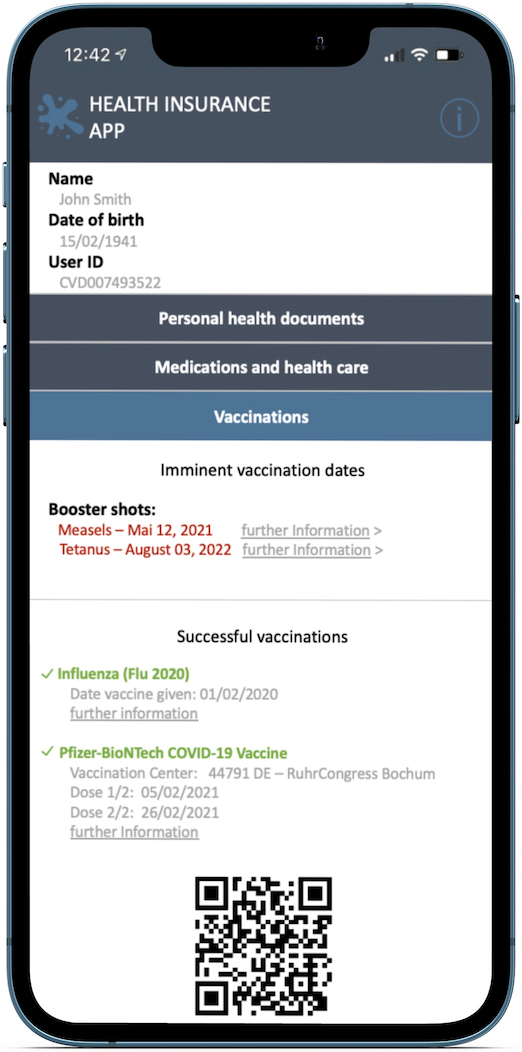}
         \caption{Health Insur. App (C4)}
         \label{fig:mockup-c4}
     \end{subfigure}
     \begin{subfigure}[b]{0.245\textwidth}
         \centering
         \includegraphics[width=\textwidth]{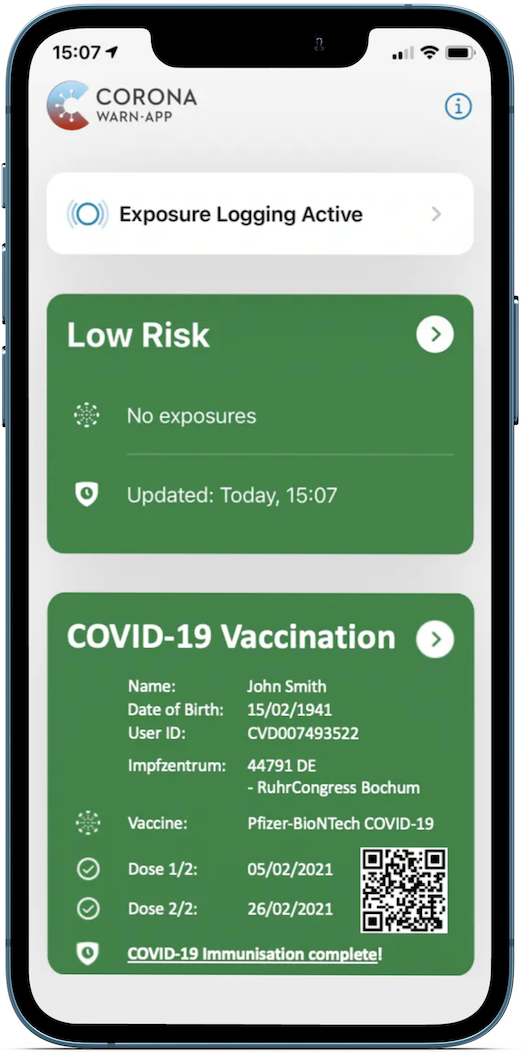}
         \caption{Corona-Warn-App (C5)}
         \label{fig:mockup-c5}
     \end{subfigure}

\caption{ Mock-ups of vaccination certificates, one of which was shown to each participants in the questionnaire~\label{fig:mockups}} 
\end{figure*}

\subsection{Variants of Vaccination Certificates (Between-Subject Conditions)}

We designed five between-subject \emph{conditions}, representing different variants of vaccination certificates.
Each condition within our survey consists of a short descriptive text supplemented by a visualized mock-up design (cf.\ Figure \ref{fig:mockups}) of the respective vaccination certificate. 
All mock-ups are based on currently used vaccination certificates or vaccination certificates planned to be issued in the near future (see Section~\ref{sec:vaxcert}).
Across all conditions, the vaccination certificates contain at least the person's \emph{name} and \emph{date of birth}, as well as the \emph{vaccinated vaccine} and \emph{date of the first and second vaccination}.
We tested two paper-based variants (C1,C2) and three digital app variants (C3--C5). In the following, all five variants are explained in detail:

\begin{compactenum}[(C1)]
\item \emph{Covid Certificate.} A simple piece of credit card-sized paper confirming the vaccination against the coronavirus. This form of documentation is currently issued by default in various countries such as the UK and USA~\cite{us21:vaccination-record-us, cdc_ukproofofvax_2020}. 
\item \emph{WHO Certificate.} An entry in the personal \emph{International Certificates of Vaccination} document standardized by the World Health Organization, which is already used by many countries to verify the vaccination status of various vaccinations~\cite{internationalcertificatesofvax_2021}.

\item \emph{Covid Certificate App.} A dedicated Corona Vaccination App simply displaying the vaccination status. This app is based on Israel's \emph{Green Pass} and the currently developed German \emph{CovPass} app which can be used throughout Europe~\cite{covpass_2021, israel21:digital-green-pass, eu21:digital-green-certificate}.

\item \emph{Health Insurance App.} A general Health Insurance Company App which is gradually introduced in Germany and further European countries to digitize the patient's record, also including vaccination records. This app is issued by health insurance companies in accordance with the \emph{Electronic Patient Record}. The app will enable the insured to view all personal sensitive health data centrally and make it easier for hospitals and doctors to share previous illnesses and patient records~\cite{personalelectronichealth_2021, federalministryepa_2020}.

\item \emph{Corona-Warn-App (CWA).} An additional entry within the CWA, which is the official contact tracing app in Germany~\cite{sap_coronawarnapp_2020}.
In addition to digital contact tracing, more features (\eg, proof of negative test results and QR-code-based check-in for events) have been continuously integrated into the app.
There is prospect that also the EU-issued \emph{Digital Green Certificate} for vaccination will be integrated into the app when available~\cite{sap_coronawarnapp_2020}.
\end{compactenum}

These five conditions can be categorized along two dimensions 
\begin{inparaenum}[(i)]
\item \textit{Paper-based} vs. \textit{app-based} solutions, which differ in the medium how the information is transmitted, and 
\item \textit{specific-purpose} vs. \textit{general-purpose} solutions, where the former are designed specifically to verify only the COVID-19 vaccination status, and the latter contain more information, \eg, contact tracing, a participant's digital health record, or various vaccinations in general. 
\end{inparaenum}

The Health Insurance App (C4) and the Corona-Warn-App (C5) are similar in that both represent integrated app-based solutions. We included both in our survey as they serve different aspects: 
\begin{inparaenum}[(i)]
\item Apps issued by health insurance companies are presumably less spread in Germany but reflect a more general setting with the potential for broader insights and adoption also in future scenarios~\cite{heisesechealthapps_2020, federalministryepa_2020, personalelectronichealth_2021}.
\item In Germany, the Corona-Warn-App has found broad adoption (approx. 28 million downloads) and therefore, integrating the COVID-19 vaccination documentation would instantly serve a huge number of users~\cite{rki_downloadzahlen_2020}.
At the same time, this integration would be limited to the purpose of the corona vaccination. 
\end{inparaenum}

\subsection{Use Case (Within-Subject Factors)}
\label{sec:method-use-cases}
A vaccination certificate has different uses.
Our study considers three within-subject \emph{use cases} (U1--U3). A specific condition, i.\,e. variant of vaccination certificate, within our survey, is always shown to a participant including all three use cases. Each use case is shown in chronologically equal order:
\begin{compactenum}[(U1)]
\item \emph{Documentation.} In the first use case, it is assumed that the respective vaccination certificate is used for documentation-purposes only.
\item \emph{Vaccination Certificate for Privileges -- Limited vaccine.} Within this use case, the actual situation -- not only in Germany -- is represented: Vaccine is only available to a limited extent, this means, not everyone has been offered a vaccination. Vaccinated citizens might have certain benefits, such as, no negative test result is necessary or international travel for private purposes is possible again~\cite{gerlatestcovidupdates_2020, germanynotenoughvaccine_2021}.
\item \emph{Vaccination Certificate for Privileges -- Vaccination offer for everyone.} This use case is similar to use case U2 but in this use case there is enough vaccine for everyone.
\end{compactenum}

Our use cases represent the situation in Germany at the time of the study (use case U1) as well as potential future use cases (U2,U3) associated with privileges, especially considering the facts that only limited vaccine was available (U2) but there will be vaccine for everyone in the future (U3). These use cases become highly relevant as U2 is more or less the current situation (late May 2021 in Germany).

\subsection{Questionnaire}
Next we outline the structure of our questionnaire. The complete questionnaire including all questions can be found in Appendix~\ref{sec:questionnaire}.
\paragraph*{Introduction}
First, we introduced the purpose of our study, namely to evaluate how a potential vaccination confirmation might look like for the recently started coronavirus vaccinations.
Moreover, we provided information about data collection and processing, noting that questions about the participant's health are asked, and asked participants for consent to proceed with the study.
\paragraph*{Smartphone Use and Experience With Coronavirus}
To get insights into participants' smartphone use and  handle of health data we asked whether they possess a smartphone~(Q1) and use an app or smartwatch to monitor their health data or sport activities~(Q2). Additionally they were asked questions about using the popular German Corona-Warn-App for digital contact tracing~(Q3) and previous experiences with the coronavirus, especially their concerns related to become infected with the coronavirus~(Q4--Q7).
\paragraph*{General Questions About Vaccinations} 
Next, we asked them general questions about vaccinations, including whether they possess the yellow vaccination certificate and if they received the recommended vaccinations~(Q8--Q11). 
We further asked several questions to capture participants' attitudes towards coronavirus vaccinations for themselves~(Q12, Q13) and for others~(Q14, Q15). 

\paragraph*{Vaccination Certificates}
In the main part of the study, we investigate the willingness-to-use and perceived utility of the various variants of vaccination certificates. 
We start with a brief summary of the current state of the pandemic and the state of the vaccination effort and illustrate the proof of vaccination for the participant's condition, comprising a short descriptive text and a mock-up.
Subsequently, we cycle through the three use cases~(U1--U3), always shown in the same order.

\begin{compactenum}[(U1)]
\item For U1, we ask how likely participants would use the presented certificate~(Q18), and further ask them to provide an open-ended response to explain their decision~(Q19--Q21).
\item For U2, we present \num{11} activities and ask participants which of these activities the certificate should be required for~(Q23) and about their willingness to use the certificate for one or more of these purposes~(Q24).
We also ask participants to rate the utility and effort to use the vaccination certificate~(Q25, Q26).
\item For U3, we again ask them about their willingness to use the certificate (Q29, under the last use case of enough vaccine being available for everyone) and reasons for their decision in a open-ended follow-up question~(Q30). 
\end{compactenum}

\noindent
We conclude this part of the questionnaire with three more general questions, asking participants which variant of vaccination certificate they generally prefer (this time, offering each participant to choose from paper- and app-based variants, if any), mandatory vaccinations and whether they perceived the pandemic as a burden~(Q32--Q34). 
\paragraph*{Disposition to Privacy}
Vaccination status, other personal information included in the certificates, and health-related information included for example in (C2) and (C4) are highly sensitive personal data, with the potential to raise privacy concerns.  
To get insights into the participants' privacy attitudes we use prior existing scales \cite{Li_2014_privacy, Chen_2012_PrivacyApps} to form two~\emph{Privacy Disposition Scales} measuring participants \emph{general} disposition to privacy (Q35, \emph{PDS-general}) with one version being specifically tailored to a mobile app context (Q36, \emph{PDS-app}).
Both scales consist of three items each, and we added a fourth question covering especially health data. 
We translated the items into German to avoid distortions due to misunderstanding.

\subsection{Pilot Study}
We tested the survey design in a pilot study with \num{150} participants recruited through convenience sampling via a university related mailing-list to get first insights and to further improve our study design.
Based on the results of the pilot study we inserted more information about the use cases on separate pages in our main study, so the differences of our use cases became more visible and thus lead to more meaningful results. Moreover, we added the questions which vaccination certificate they generally prefer (Q32).

Other than that, we used our pilot study to validate the privacy questions. 
Both privacy scales, PDS-general and PDS-app, showed good to perfect internal consistency scores (Cronbach's $\alpha = $ .78 and .94). Therefore, we proceeded with the main study as planned. 

\subsection{Data Collection and Sample}

\begin{table}[tp]
    \centering
    \begin{tabular}{lrrrrrr@{\hspace{0.15cm}}rr}
    \toprule
     & C1 & C2 & C3 & C4 & C5 & \multicolumn{2}{c}{Total} & Target \\
    \midrule
    Female & \num{60} & \num{64} & \num{62} & \num{60} & \num{63} & \num{309} & \SI{51.6}{\percent} & \SI{51.2}{\percent} \\
    Male & \num{60} & \num{55} & \num{57} & \num{59} & \num{58} & \num{289} & \SI{48.2}{\percent} & \SI{48.8}{\percent} \\
    Non-binary & \num{0} & \num{1} & \num{0} & \num{0} & \num{0} & \num{1} & \SI{0.2}{\percent} & -- \\
    \midrule
    \num{18}--\num{24}  & \num{11} & \num{10} & \num{10} & \num{9} & \num{8} & \num{48} & \SI{8.0}{\percent} & \SI{11.2}{\percent} \\
    \num{25}--\num{34}  & \num{18} & \num{18} & \num{19} & \num{18} & \num{20} & \num{93} & \SI{15.5}{\percent} & \SI{15.2}{\percent} \\
    \num{35}--\num{44}  & \num{20} & \num{20} & \num{19} & \num{19} & \num{18} & \num{96} & \SI{16.0}{\percent}  & \SI{14.4}{\percent} \\
    \num{45}--\num{54}  & \num{24} & \num{24} & \num{21} & \num{22} & \num{22} & \num{113} & \SI{18.9}{\percent} & \SI{18.4}{\percent}\\
    \num{55}--\num{65}  & \num{23} & \num{21} & \num{25} & \num{24} & \num{24} & \num{117} & \SI{19.5}{\percent} & \SI{18.4}{\percent}\\
    \num{66}--\num{90}  & \num{24} & \num{27} & \num{25} & \num{27} & \num{29} & \num{132} & \SI{22.0}{\percent} & \SI{22.4}{\percent}\\
    \bottomrule
    \end{tabular}
    \caption{Gender and age distribution of participants across conditions (C1--C5). Target quotas for representativeness were matched with all deviations smaller than \SI{3.2}{\percent}.}
    \label{tab:demographics}
\end{table}

Participants were recruited by the panel provider Lightspeed Research (Kantar) from an online panel and sampled representatively for the population in Germany wrt.\ gender, age and education.  
The demographics of our sample are listed in Table~\ref{tab:demographics}.
Deviation from the target demographics is smaller than \SI{3.2}{\percent} for all targets.
Within our sample, there are 309 female participants (\SI{51.6}{\percent}), 290 male participants (\SI{48.2}{\percent}), and 1~non-binary participant.
The average age was \num{49.6} years and \SI{85}{\percent} of the participants do not have an education in, nor do work in, the field of computer science, computer engineering or IT. 
Participants took on average about \num{12}~minutes to complete the online survey.

\subsection{Research Ethics}
Our department does not have an institutional review board. Instead, our study followed best practices of human subject research and data protection guidelines.
To  minimize any potential adverse effects from the study we followed the ethical principles laid out in the Belmont report~\cite{belmontreport_2018}, specifically we sought informed consent at the beginning of the study and participants were informed that they could withdraw from the study without any negative consequences at any time. 
Kantar, our panel provider, has committed itself to follow the ICC/ESOMAR code of conduct~\cite{ICCESOMAR_Code_2021}.

%% file: sections/4-results.tex
\section{Results}

In the following section we present  our results.  Our main results concern the participants' \emph{willingness to use} various forms of vaccination certificates and the factors that influence it. Further results concern the \emph{perceived effort} and the \emph{perceived utility} of the different types of vaccination certificates. 

\begin{figure*}[ht]
\centering
\includegraphics[width=\textwidth]{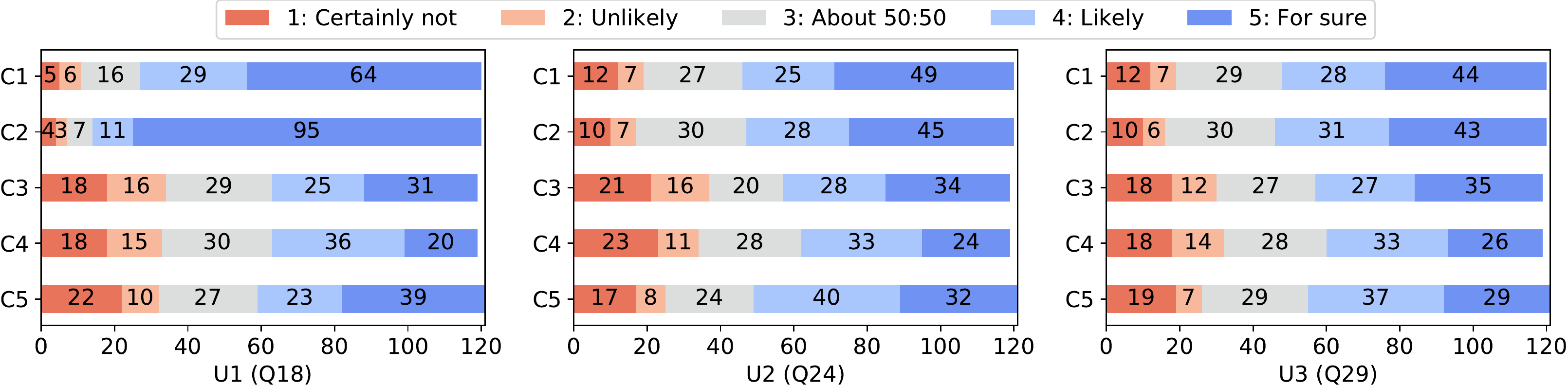}
\caption{Overview of participants' willingness to use vaccination documentations across the five conditions~(C1--C5) and three use cases~(U1--U3). Responses were collected on equidistant five-point scales.} 
\label{fig:willingness-to-use-distributions}
\end{figure*}

\subsection{Willingness to Use Vaccination Certificates}

Using vaccination certificates is voluntary in Germany where we conducted our survey.  
Thus, willingness to use such certificates plays an important role for their adoption. 
In the first step of our evaluation, we analyze the responses to questions about participants' willingness to use vaccination certificates across the five conditions and three use cases.
The questions were answered on equidistant five-point scales by Rohrmann~ \cite{rohrmann07:verbal-qualifiers-rating, rohrmann78:empirische-studien-entwicklung}. 
Higher response values represent more positive responses, \ie, a higher willingness to use the described certificate form.
We compare the responses using analysis of variance (ANOVA) and post-hoc pairwise tests with Bonferroni correction. 
When requirements for the analysis of variance such as homogeneity of variances were violated we used Welch's ANOVA instead. 

Subsequently, we analyze which factors impact participants' willingness to use a vaccination certificate. 
We therefore apply a multiple linear regression for both continuous and categorical variables to analyze covariances.
We perform one covariance analysis per use case for U1 and U2 with the willingness to use (Q18 and Q24) as dependent variables. 

When determining significance for all results ($\alpha = $~\SI{5}{\percent}), we used Bonferroni-Holm corrected alpha values to ensure multiple testing correction. For our results we indicate significance-levels with stars (*$p<.05$, **$p<.01$, ***$p<.001$). 

\subsubsection{Differences Between Conditions and Use Cases}
\label{sec:wtouse-differences}
Figure~\ref{fig:willingness-to-use-distributions} illustrates the distribution of responses to questions about willingness to use the different types of vaccination certificates separately for each use case.
Accordingly, Table~\ref{tab:willingness-to-use-mean} lists the mean response values for these questions.
Mean responses were generally rather positive across all conditions and use cases. Most strikingly, we observe the highest willingness to use for the standardized paper-based international vaccination certificate when it is only used for documentation purposes (C2 $\times$ U1, $mean=4.58$).
In this scenario, \num{106} out of \num{120} participants provided a positive answer.

For the documentation use case (U1) we found pairwise statistically highly significant differences ($p<.01$) between each of C1 and C2 compared to C3, C4, and C5.
This means that both paper-based certificates reach higher preference rates than each of their app-based counterparts.
We also observe tendencies to higher response values for C1 and C2 compared to the other conditions in the other two use cases (U2 and U3).
However, these differences are only significant for U2 ($p<.05$) and not statistical significant for U3. 
Additionally, we observe significantly higher willingness to use values for the documentation use case (U1) compared to both certificate uses cases (U2 and U3) across all conditions (\cf~Table~\ref{tab:willingness-to-use-mean}, \emph{Overall}). 
Since differences between scenarios can be mainly observed between paper-based (C1,C2) and app-based (C3--C5) conditions, we summarize those groups of conditions for our subsequent analyses.

\begin{table*}[tb]
\centering
\caption{Mean response values for questions about willingness to use vaccination certificates across conditions~(C1--C5) and use cases~(U1--U3). For each use case, groups indicate significant differences between conditions. Willingness to use differs significantly between two conditions, when they are not part of the same group (\ie, A or B). \label{tab:willingness-to-use-mean}}

\begin{tabular}{l@{\hspace{0.1cm}}lclll@{\hspace{1cm}}clll@{\hspace{1cm}}clll}
\toprule

&& \multicolumn{4}{c}{Use Case U1} & \multicolumn{4}{c}{Use Case U2} & \multicolumn{4}{c}{Use Case U3}\\
\multicolumn{2}{l}{Condition} & mean $\pm$ sd & \multicolumn{3}{l}{group} & mean $\pm$ sd & \multicolumn{3}{l}{group} & mean $\pm$ sd & \multicolumn{3}{c}{group} \\

\midrule

C1 & Covid Cert. & \num{4.18} $\pm$ \num{1.11} & A & & & \num{3.77} $\pm$  \num{1.31} & A & & & \num{3.71} $\pm$   \num{1.29} & A & &   \\
C2 & WHO Cert. & \num{4.58} $\pm$ \num{0.96} & A & & &  \num{3.76} $\pm$  \num{1.25} & A & & & \num{3.76} $\pm$   \num{1.23} & A & &\\
C3 & Cert. App & \num{3.29} $\pm$ \num{1.39} & & B & & \num{3.32} $\pm$  \num{1.46} & A & B & & \num{3.41} $\pm$   \num{1.40} & A & & \\
C4 & Insur. App &  \num{3.21} $\pm$ \num{1.29} & & B & & \num{3.20} $\pm$  \num{1.39} & & B & & \num{3.29} $\pm$   \num{1.34} & A & & \\ 
C5 & CW App & \num{3.39} $\pm$ \num{1.47} & & B & & \num{3.51} $\pm$  \num{1.33} & A & B & & \num{3.41} $\pm$  \num{1.34}  & A & &\\
\midrule
& Overall &  \num{3.71} $\pm$ \num{1.37} &&&& \num{3.51} $\pm$ \num{1.36} &&&& \num{3.52} $\pm$ \num{1.33}\\
\bottomrule
\end{tabular}

\end{table*}

After all use cases, we explicitly asked participants if they preferred an app-based or paper-based vaccination certificate. \SI{44}{\percent} of participants would rather use a paper-based variant, compared to \SI{37}{\percent} rather using an app-based variant. %(\cf~Table~\ref{tab:totalvariants}).
Around \SI{12.5}{\percent} were undecided if they would use any certificate and \SI{6.5}{\percent} of participants would not use any vaccination certificate.
As \SI{40}{\percent} of participants were in a paper-based condition (C1 or C2) and \SI{60}{\percent} of participants were in an app-based condition (C3--C5), these results do not solely reflect the condition participants were in, but rather show their general opinion.  

\subsubsection{Predictors for Differences} 

\begin{table}[tb]
\centering
\caption{Covariance analysis for participants' willingness to use vaccination certificates.  A positive estimate indicates participants’ willingness to use the certificate being higher compared to the factor’s baseline. Significance-levels are indicated with stars (*$p<.05$, **$p<.01$, ***$p<.001$).  \label{tab:estimates}}
\begin{tabular}{lrr}
\toprule
Independent variables & U1 (Q18) & U2 (Q24) \\
\midrule
\multicolumn{3}{l}{\emph{Condition (baseline: paper)}} \\
App & \num{-.96}*** & \num{-.32}***  \\
\midrule
\multicolumn{3}{l}{\emph{[Q3]: Uses German Corona-Warn-App (baseline: no)}} \\
Yes & \num{.57}*** & \num{.47}*** \\
Not anymore & \num{.24} & \num{.18} \\
Not sure & \num{-.41} & \num{.55} \\
\midrule
\multicolumn{3}{l}{\emph{[Q4/Q5]: Coronavirus Infection (baseline: no)}} \\
Yes & \num{.04} & \num{.08} \\
\midrule
$[$Q6/Q7$]$: Worries about Coronavirus & \num{.24}*** & \num{.20}***  \\
\midrule
\multicolumn{3}{l}{\emph{[Q12/Q13]: Vaccination Willingness (baseline: no)}} \\
Yes & \num{.56}*** & \num{.70}***  \\
\midrule
\multicolumn{3}{l}{\emph{[Q33]: Vaccination Obligation (baseline: Unsure)}} \\
Yes & \num{.11} & \num{.14}  \\
No & \num{-.24}* & \num{-.39}*** \\
\midrule
$[$Q34$]$: Pandemic as Burden & \num{.01} & \num{.06} \\
\midrule
$[$Q35$]$: Disposition to Privacy & \num{-.19}*** & \num{-.13}* \\
\midrule
\multicolumn{3}{l}{\emph{Gender (baseline: Male)}} \\
Female & \num{-.06} & \num{-.11} \\
Non-binary & \num{.54} & \num{1.19} \\
\midrule
\multicolumn{3}{l}{\emph{Education (baseline:High school degree)}} \\
No graduation & \num{-.69} & \num{.47} \\
Still in school & \num{-.50} & \num{.001} \\
Practical training & \num{-.05} & \num{.02} \\
University degree & \num{.09} & \num{.03} \\

\bottomrule
\end{tabular}
\end{table}

We determine the factors that explain the differences between the scenarios reported in Section~\ref{sec:wtouse-differences}, particularly between the paper-based and app-based conditions in the documentation use case (U1) and the vaccination certificate for privilege use case (U2). 
To this end, we create a new \emph{Condition} variable with two levels \emph{paper-based} (reflecting C1 and C2) and \emph{app-based} (C3--C5). We also group other related factors for the analysis: For Q6 and Q7 we group the answers into a score showing ``worries about Coronavirus'', similar to our approach with the Privacy Disposition Score. For Q4 and Q5 we create a new factor ``Coronavirus Infection'' indicating ``yes'' for participants answering ``yes'' to at least one of the questions and ``no'' for participants answering ``unsure'' or ``no'' to both of these questions. We proceed the same way for Q12 and Q13, creating the new factor ``Vaccination Willingness''.

We conducted covariance analyses examining the willingness to use as outcome variable (Q18 for U1 and Q24 for U2) for a set of factors as listed in Table~\ref{tab:estimates}.
Reported estimates indicate influence on participants' willingness to use vaccination certificates.
Positive values indicate higher willingness to use the respective certificate, negative values indicate lower willingness.

We find the \emph{condition} (paper-based vs. app-based), attitudes towards \emph{vaccination obligation} (Q33), and the participants' \emph{disposition to privacy} (PDS-general, Q35) being significantly inhabitant factors for the willingness to use vaccination documentation and certificates. 
Participants' are less willing to use an app-based vaccination documentation/certificate than a paper-based version. 
This effect is stronger for U1 (\num{-0.96}) than for U2 (\num{-0.32}). 
Participants with higher disposition of privacy are also less willing to use a vaccination documentation or certificate. 
Again, the effect is slightly stronger for U1 than for U2.

Being more \emph{worried about Coronavirus Infection} (Q6/7) and having a positive attitude towards vaccination (Q12/13) positively influence the willingness to use for use cases U1 and U2.
Participants being worried about the coronavirus are more willing to use a vaccination documentation and certificate, with a slightly higher estimate for U1. Participants who are willing to get vaccinated or are already vaccinated are more willing to use a vaccination documentation and certificate than participants who do not want to get vaccinated, with a higher estimate for U2. 
Additionally, participants using the German \emph{Corona-Warn-App} (Q3) are more willing to use vaccination documentation and certificates, with a slightly higher estimate for U1.

\subsubsection{Analysis of Significant Factors}
For a deeper understanding of factors that affect willingness to use vaccination certificates, we also take a closer look into participants' open-ended responses.
We asked for reasons why or why not participants would use certificates in use cases U1 (Q19--Q21) and U3 (Q30), depending on their responses to Q18 and Q29, respectively.
To evaluate these open-ended responses, we followed an iterative coding procedure.
Two researchers independently assigned codes to the responses to identify and group common themes. 
Each response could be assigned multiple codes.
The two coders then discussed their individual codings, agreed on a final coding scheme and aligned their individual codes to this scheme, finalized by a mutual validation of codings. 

Table~\ref{tab:quali-codes} summarizes the numbers of participants in each group (\ie, condition and willingness-to-use), along with the number of open-ended responses and codes assigned to these responses.
For willingness-to-use, we group participants based on their responses to Q18 (and Q29, respectively).
Participants who provided the two lowest response levels on the five-point response scale are labeled \emph{No}, the two highest response labels are assigned \emph{Yes} and the neutral response is labeled \emph{Undecided}.

\begin{table}[tbp]
\centering
\caption{Coding Statistics. Numbers of total participants (\emph{n}), responses (\emph{resp}) and assigned codes, grouped by condition and willingness to use the certificate for two use cases (Q18/29). \label{tab:quali-codes}}
\begin{tabular}{llrrrrrr}
\toprule
&& \multicolumn{3}{c}{U1 (Q19--Q21)} & \multicolumn{3}{c}{U3 (Q30)} \\
Cond.& Q18/29 & n & resp & codes & n & resp & codes \\

\midrule

\textbf{Paper} & No & 18 & 13 & 14 & 35 & 31 & 34\\
& Undec. & 23 & 15 & 15 & 59 & 28 & 30\\
 & Yes & 199 & 184 & 204 & 146 & 126 & 137\\
\textbf{App} & No & 99 & 91 & 106 & 88 & 82 & 92\\
& Undec. & 86 & 82 & 87 & 84 & 63 & 65\\
 & Yes & 174 & 148 & 170 & 187 & 156 & 172\\

\bottomrule
\end{tabular}
\end{table}

\paragraph*{Privacy Concerns and Disposition}

When asked about reasons for not using a vaccination certificate for documentation (U1, Q19), about one third of responses in the app-based conditions (30 out of 91 responses) referred to privacy concerns.
The majority of participants mentioned general concerns about privacy (P407: \emph{``I am concerned about data protection''}) or referred to specifics of smartphone privacy (P467: \emph{``I do not need any more data collection on my smartphone''}, P517: \emph{``I do not want to be tracked''}).
Additionally, a few responses explicitly referred to privacy concerns related to health data, \ie, sensitive information (P255: \emph{``Because I do not want any health data on my phone''}, P509: \emph{``I do not want my health data to be stored on my phone''}).
Privacy being an important factor influencing willingness to use a vaccination certificate is also reflected by participants who were undecided in their willingness to use the certificate. 
Among those participants \num{28} responses (out of \num{82} responses we received) referred to privacy concerns (P355: \emph{``Because this is another app that secretly reads data on my smartphone''}, P475: \emph{``I am not sure if I trust smartphone apps in terms of privacy [...]}).

Among the app-based conditions, we observe the highest number of privacy references in C4 (Health Insurance App) and the fewest in C3 (Covid Certificate App).
In the paper-based conditions, no responses referred to such concerns about privacy or protection of sensitive information, independent of participants' willingness to use the certificate.
Interestingly, the number of responses referring to privacy is lower in use case U3, \ie, when the certificate is used as a proof that can be used for specific purposes.
Among participants who are not willing to use the certificate in this scenario, \num{21} (out of \num{82}) responses refer to privacy, and only \num{6} (out of \num{63}) of those who are undecided.

These observations confirm the results of our covariance analysis, which shows a \emph{strongly} significant negative influence of participants privacy disposition in use case U1, and a smaller significant influence for U3~(\cf~Table~\ref{tab:estimates}).

In this context, we also take a closer look into participants' privacy attitudes, \ie, their Privacy Disposition Score in general (PDS-general, Q35) and Privcacy Disposition Score related to apps (PDS-app, Q36).
Overall, PDS-general ($mean=3.26, sd=0.81$) is slightly higher than PDS-app ($mean=3.05, sd=1.22$), along with larger standard-deviations for PDS-app compared to PDS-general. %~(\cf~Table~\ref{tab:pds}). 
We also observe moderate significant correlations for PDS-general and U1 ($r=.17$**) as well as for U2 ($r=.14$**) across all conditions. 
For PDS-app and use cases U1 and U2, we observe high and significant correlations (both $r=.47$**), showing a strong relation for the willingness to use an app-based vaccination documentation or certificate with PDS-app. 
This again highlights privacy disposition as a significant factor for the (planned) adoption of vaccination documentation and certificates, especially if these are app-based.

\paragraph*{Attitudes Towards Vaccination}
As we see in the results of the covariance analysis, participants' willingness to get vaccinated has a strong significant positive effect on their willingness to use a certificate. Additionally, their attitude towards vaccinations being mandatory has significant influence on their willingness to use vaccination certificates.

Since responses reflecting participants' general attitudes towards vaccinations do not depend on whether they were asked about paper-based or app-based certificates, we do not distinguish conditions here.
Among all \num{117} participants who were not willing to use a vaccination certificate, \num{13} (out of \num{104} who provided a response) indicated that they were not going to be vaccinated. However, responses come in different flavors, ranging from individuals who are ineligible for vaccination, \eg, for medical reasons (P35: \emph{``I must not get vaccinated.''}), to anti-vaccinationists (P10: \emph{``Because I am principally opposed to vaccinations.''}, P441: \emph{``I do not intend to get vaccinated against an unresearched disease.''}), to conspiracy theorists (P132: \emph{``No one will ever inject me that poison.''}, P249: \emph{``The corona pandemic is simply overplayed [...] Everyone will die anyway, so it's better to let it just happen.''}).

Among those participants who were undecided about their willingness to use the certificate, \num{7} (out of 97) responses indicate that participants have not finally decided about getting vaccinated.

Regarding attitudes towards mandatory vaccinations, several participants expressed that they did not appreciate a form of \emph{indirect vaccination obligation} in use case U3.
\num{13} participants who are not willing to use a certificate indicated this in their open-ended responses (P236: \emph{``They create some kind of vaccination obligation [...]''}, P526: \emph{``This way, getting vaccinated is no longer voluntary.''}).
In a similar fashion, \num{12} participants referred to \emph{discrimination} between vaccinated and not vaccinated people (P189: \emph{``All people should have the same rights and opportunities.''}, P274: \emph{``That would discriminate all people who do not want to get vaccinated, for whichever reason.''}).
On the opposite side, one participant who was willing to use their paper-based certificate also referred to vaccinations being mandatory, however, without opposing this (P8: \emph{``This way, the vaccination will become mandatory. I do not have a problem with that.''}).

%\vspace{-0.5cm}
\paragraph*{Worries about Coronavirus}
We identified \num{51}~responses to Q30 (U3), in which participants referred to their intent to \emph{protect themselves or others} from the coronavirus as a justification to use their vaccination certificate for specific activities~(Q30) (P241: \emph{``for my own safety''}, P138 \emph{``to protect myself and others''}) across all five conditions (\num{35} in paper-based conditions and \num{16} in app-based conditions).
Eleven participants (paper: \num{9}, app: \num{2}) mentioned \emph{fighting the pandemic} a reason to use their certificate (P488: \emph{``Help containing the corona pandemic''}).
Among participants who are not willing to use any form of certificate we did not identify references to reasoning about health protection or fighting the pandemic.
This is in line with the finding that willingness to use a vaccination certificate is significantly higher when participants are worried about the coronavirus, \eg, worried that they or people close to them could catch a coronavirus infection (\cf~Table~\ref{tab:estimates}).

\subsubsection {Certificate Purposes}
\begin{figure}[t]
\centering
\includegraphics[width=0.5\columnwidth]{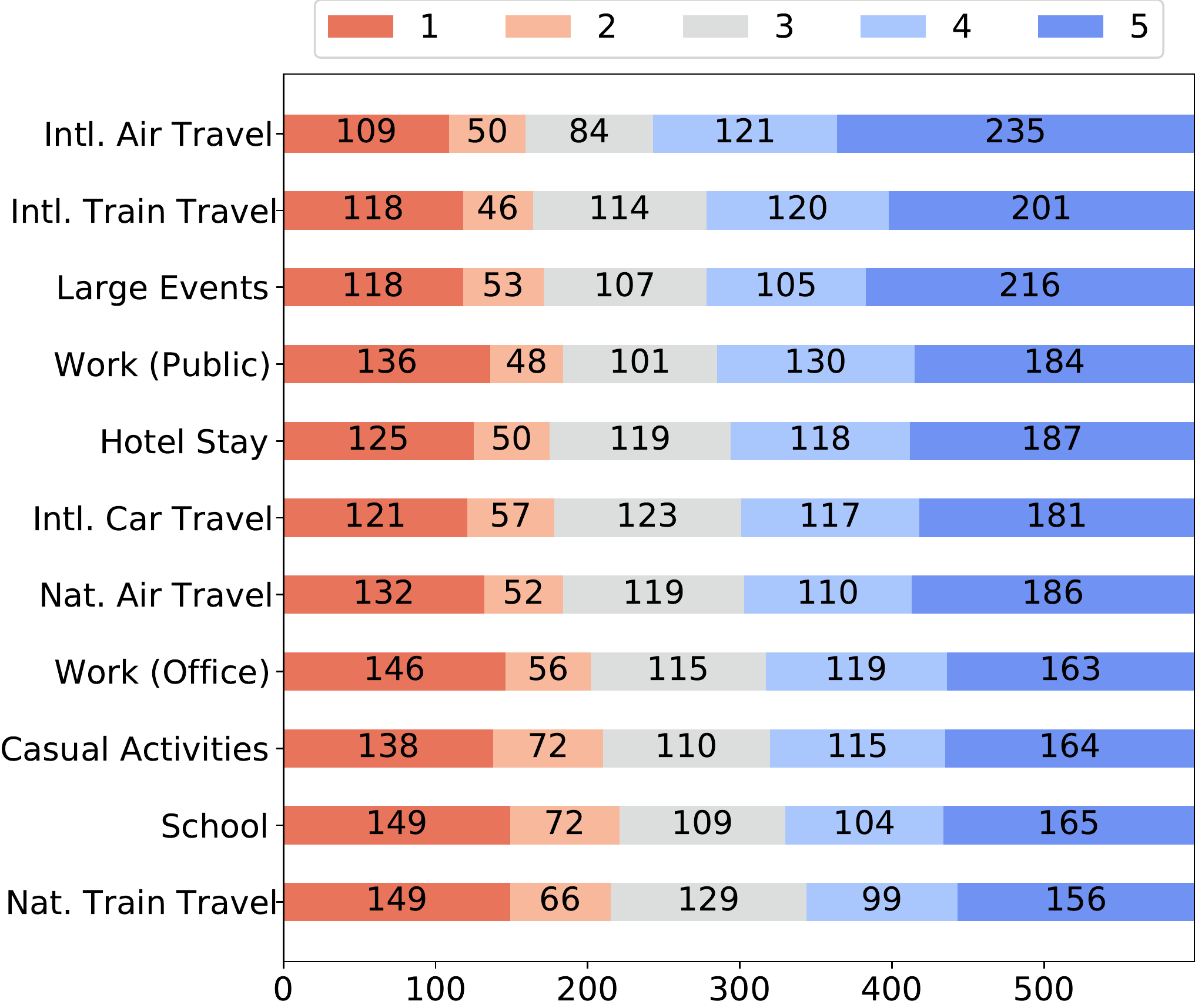}
\caption{Overview of participants' willingness to use vaccination certificates as a requirement for specific activities (Q28). Responses were collected on equidistant five-point scales.} 
\label{fig:q28-top}
\end{figure}

We conclude our analysis of willingness to use vaccination certificates with a look into participants' attitudes towards specific purposes, \ie, whether they would use the certificate for them. 
Within the scenario for use case U3, we asked participants what purposes they would use the certificate for (Q28).
In total, we asked for \num{11} purposes covering a broad range of business, leisure, or travel activities.
Participants indicated their willingness to use their vaccination certificate on five-point scales for each purpose.
While we only observe minor differences among the five conditions for each purpose, we find tendencies indicating that participants are more willing to use their vaccination certificate for more extraordinary purposes.
Across all conditions, \num{356} out of \num{599} participants provided positive responses for \emph{International Air Travel}, compared to \num{255} positive responses for \emph{National Train Travel}, which also includes local public transport.
Similarly, using the certificate to attend \emph{Large Events} such as sports events or concerts finds broader acceptance than more \emph{Casual Activities} such as going to a restaurant.
The distribution of responses to all purposes we considered is illustrated in Figure~\ref{fig:q28-top}. 

\subsection{Effort and Utility of Vaccination Certificates} \label{sec:effortandutility}

\begin{figure*}[ht]
\centering
     \begin{subfigure}[h]{0.49\textwidth}
         \centering
        \includegraphics[width=\columnwidth]{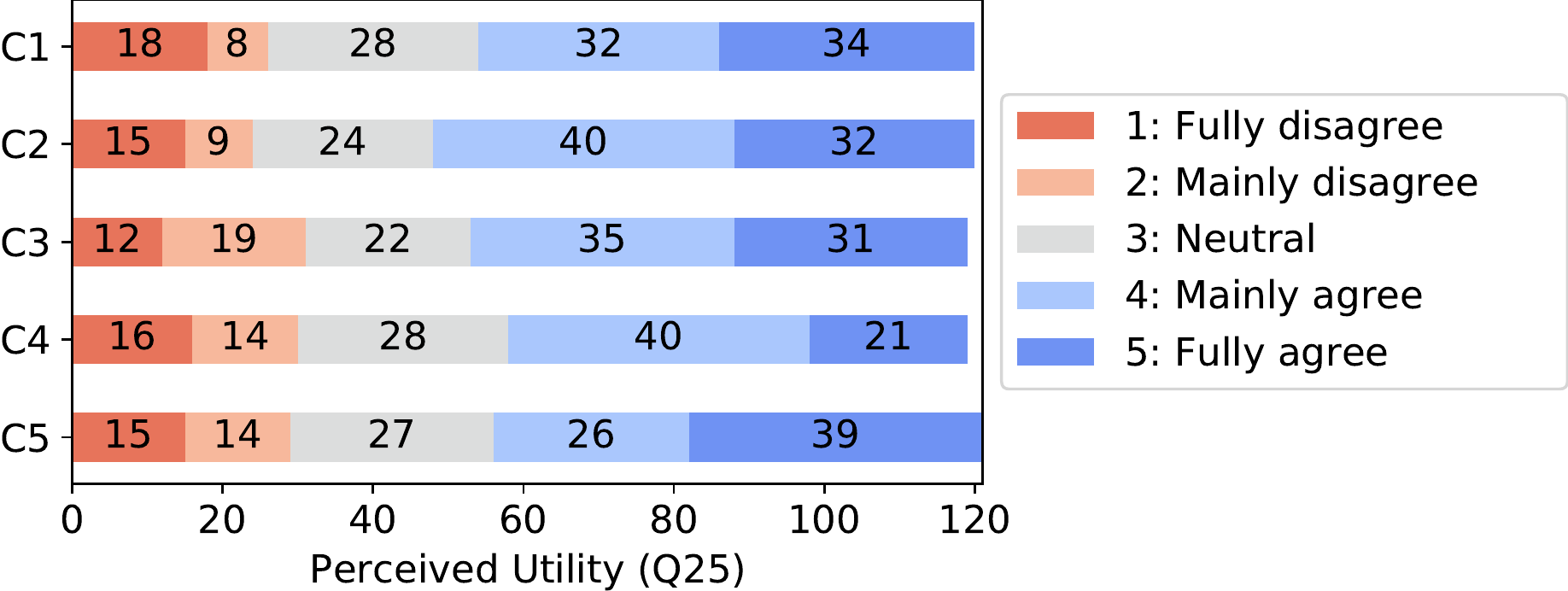}
     \end{subfigure}
     \begin{subfigure}[h]{0.49\textwidth}
        \centering
        \includegraphics[width=\columnwidth]{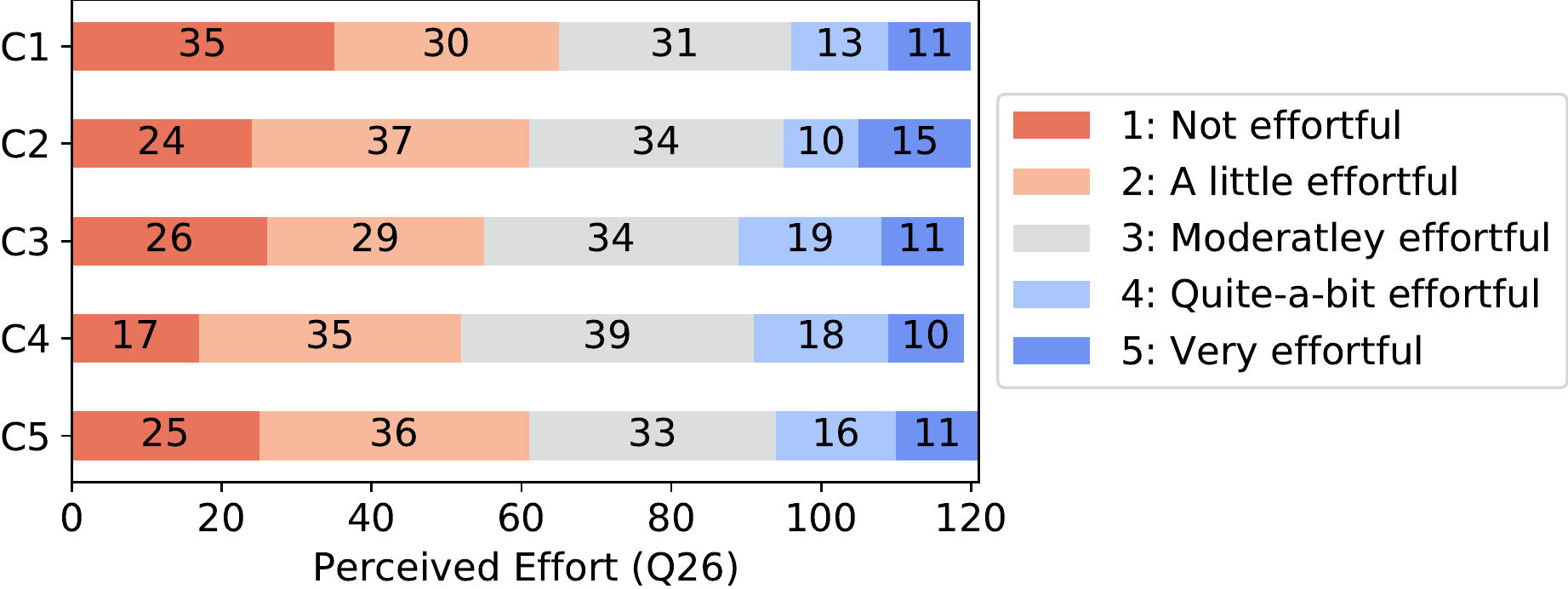}
     \end{subfigure}
     
\caption{Overview of participants’ effort and utility to use vaccination certificates across the five conditions (C1–C5) in use case U2. Responses were collected on equidistant five-point scales.~\label{fig:effort-utility-distributions}} 

\end{figure*}

In use case U2, we asked participants to rate the utility~(Q25) and effort~(Q26) of their vaccination certificate. 
Larger values express higher utility and lower values show less effort. 
The results in Table~\ref{tab:effort-utility} and Figure~\ref{fig:effort-utility-distributions} show that all certificates are viewed as rather useful ($mean$ $>$ \num{3}) as well as rather easy to use (effort $mean$ $<$ 3). 
We only observed slight differences between the conditions, with none of them being significant. 
There is a strong positive correlation~($r=.76$) between utility and willingness to use~(Q24), and a negative correlation~($r=-.46$) between effort and willingness to use.
Finally, effort and utility correlate negatively~($r=-.39$).

\begin{table}[b]
\centering
\caption{Effort and Utility of the vaccination certificates (U2- mean values and standard-deviation). \label{tab:effort-utility}}
\begin{tabular}{lcc}
\toprule
Condition &  Utility (Q25) & Effort (Q26)  \\
& [mean $\pm$ sd] & [mean $\pm$ sd] \\
\midrule

\textbf{C1} & \num{3.47} $\pm$ \num{1.37} & \num{2.46} $\pm$ \num{1.27} \\
\textbf{C2} & \num{3.54} $\pm$ \num{1.30} & \num{2.62} $\pm$ \num{1.25} \\
\textbf{C3} & \num{3.45} $\pm$ \num{1.31} & \num{2.66} $\pm$ \num{1.24} \\
\textbf{C4} & \num{3.30} $\pm$ \num{1.27} & \num{2.74} $\pm$ \num{1.14} \\
\textbf{C5} & \num{3.50} $\pm$ \num{1.37} & \num{2.60} $\pm$ \num{1.21} \\
\midrule
\textbf{Overall} & \num{3.45} $\pm$ \num{1.32} & \num{2.62} $\pm$ \num{1.22} \\
\bottomrule
\end{tabular}
\end{table}

Within the open-ended responses collected \wrt willingness to use we also gained insights into participants' perceived \emph{utility} of the certificates.
Since willingness to use a certificate and its utility are highly correlated, it is plausible that utility aspects were named reasons for why participants would use a certificate, both for documentation (U1), and also as a certificate as requirement for specific activities (U3).
For documentation purposes (U1), \num{7} participants referred to utility as a reason to use the certificate in the paper-based condition, and \num{48} participants referred to it in the app-based condition.
Whereas utility was mostly attributed to the certificates easy-to-carry nature (P28: \emph{``I can carry this with me in my wallet.''}, P76: \emph{``[...] simple, handy''}) in the paper-condition, participants in favor of the app-based certificate appreciated that they did not need to carry any separate item with them (P306: \emph{``On my phone I have my confirmation always with me.''}, P555: \emph{``The alternative was carrying my vaccination card at all times.''}). 

In U3, \num{3} participants referred to utility in the paper-based condition, and \num{37} in the app-based condition, providing similar arguments as for U1. 
Even though the numbers are lower in U3, this does not imply that they rated the certificate as less useful than for documentation.
In this scenario, large numbers of participants referred to being happy about using the certificate for travel or to attend events, thus putting a different focus.
However, even such answers can be considered related to utility in a broader sense.

%% file: sections/5-discussion.tex
\section{Discussion}

Our study revealed factors that impact the willingness to use vaccination certificates foremost especially designed to fight the COVID-19 pandemic. The results have implications not only for the design of vaccination certificates but also show that their utility and people's privacy disposition play an important role in their willingness to adopt vaccination apps.

\subsection{Willingness to Use Vaccination Certificates} 
Generally speaking, the willingness to use vaccination certificates is high, up to 4.58 (C2$\times$U1).
Interestingly, the use cases we considered in the survey had a significant influence on the willingness to use. Over all conditions, willingness to use was 3.73 for use case U1 (for documentation purposes only), 3.51 for use cases U2, and 3.52 for U3 (used to verify vaccination status associated with privileges). The differences in the willingness to use between use case U1 and the other two use cases are significant ($p<0.05$) -- participants are significantly more willing to use all variants of certificates in use case U1 than in both other use cases.
We suspect that this is related to the significantly more invasive intervention caused by specific privileges only given to vaccinated individuals associated with concerns about a two-class society and discrimination.

Especially among those willing to be vaccinated or already vaccinated, the willingness to use vaccination certificates is particularly high ($mean=4.07$) compared to those who refuse vaccination against the coronavirus ($mean=2.34$).
We therefore conclude that vaccination certificates will be considered by our participants as an effective means to fight the COVID-19 pandemic. 

For use case U1 (considering that the respective vaccination certificate serves only for documentation purposes) our survey indicates that people prefer paper-based variants (C1,C2) compared to app-based variants (C3--C5), these differences are highly significant. There is no significant difference between C1 and C2, that is, whether a corona-specific paper-based certificate (C1) or the more generalized yellow vaccination certificate (C2) is used.
The high willingness to use rates for the paper-based versions might be due to habits, as 76\% of our participants posses the certificate variant in C2, the yellow vaccination certificate. For the documentation purpose many participants seem to value the existing certificate as sufficient. 
In both privilege-related use cases (U2,U3) there is almost no difference between both paper-based variants.

In our qualitative analysis it was striking that many participants mentioned potential privileges when asked why they would use the respective vaccination certificate variant already for use case U1. Many participants argue they would like to be prepared once vaccination certificates become mandatory for certain purposes like traveling. This might be also due to the public debate about potential vaccination certificates and associated privileges such as traveling, attending events, or even lifting curfew, taking place in Germany during the time of our survey. 

In case vaccination certificates are mandatory for certain purposes (U2,U3), there are less significant differences between paper-based and app-based vaccination certificates. Only for use case U2 we observe significantly higher values for willingness to use for both paper-based versions compared to C4, the general health care app. This might be due to high privacy concerns that such a general app which contains more sensitive health data produce. 
We do not observe significant differences between the specific and more general paper-based versions. Especially in use cases U2 and U3 the mean values for willingness to use of both paper-based variants are similar, participants do not seem to like one of these better for those use cases (compared to use case U1). It has to be noted that none of the differences between the paper-based variants nor between app-based variants are significant.

\subsection{Privacy as a Key Factor for Willingness to Use} \label{privacyexpect}
Our results clearly indicate that there are significant (negative) correlations between the willingness to use and the participants' self-reported privacy disposition, both for the general and the app related privacy disposition. Correlations are higher for participants' willingness to use and the privacy disposition related to apps. Also the evaluated qualitative responses support these significant qualitative findings with regard to privacy concerns using vaccination apps -- around 34\% of the open answers stated privacy concerns. In none of the paper-based conditions participants mentioned privacy concerns, which are therefore most relevant for app-based certificates. This shows once again that privacy plays an essential role for the user, especially in the case of digital health apps. This is in line with related work that identified privacy concerns as limiting factors for the use and acceptance of mobile health apps \cite{wottrich_privacy_2018, zhou_barriers_2019}.
Other related work regarding the willingness to use apps to fight the current pandemic, have also shown that privacy plays an essential role when it comes to apps using sensitive personal health data \cite{utz21:apps-against-spread}. 

This is again reflected by our qualitative results, as some participants explicitly referred to privacy concerns regarding their personnel health data. Others mentioned privacy more broadly, like not wanting to be tracked, wanting their data to be protected generally, are scared of unwilling data access by the app, and do not trust their smartphone apps with data privacy. These thoughts highlight participants' deeply privacy concerns as well as their desire for privacy protection.
Privacy is also often related to anonymity, as the discussion on the CWA in Germany showed. The CWA therefore follows anonymity and a decentralized privacy friendly approach.

\subsection{Other Factors Influencing the Willingness to Use}

\paragraph*{Social Factors}
Social reasons play a significant role in the adoption of vaccination certificates -- as studies have also shown in terms of contact tracing apps~\cite{utz21:apps-against-spread}.
Social reasons here consists of protecting other people, as people who are worried about getting of a coronavirus infection (or worried about people close to them getting a coronavirus infection) have a significantly increased willingness to use vaccination certificates. Participants also frequently mentioned in their qualitative responses, that they want to use the certificates to protect themselves but also to protect others by proving their vaccination status. 
\paragraph*{Health Insurance App (C4)} \label{healthinsurance}
We observe the tendency that the willingness to use C4 (i.\,e., the universal Health Insurance App) is the lowest in all three use cases. This is the app variant that is most generalized and contains most health related data in addition to the vaccination status, such as medical records shared between hospitals and medical practices, and general registered vaccinations. We also stated in the description that this variant is made available by the health insurance company, which might have led participants to oppose this variant. This finding might also be due to the participants' privacy concerns, as this variant uses the most sensitive personal health data and therefore might trigger the most privacy concerns. This assumption is backed by our qualitative analysis, where we find that the most privacy concerns are posed for this vaccination certificate variant.
\paragraph*{Corona-Warn-App (C5)}
Another factor in regard to habits (like the paper-based preference) is the use of the German contact tracing app, the Corona-Warn-App, which was used by around \SI{40}{\percent} of the participants. We found that using the German Corona-Warn-App correlates with a higher willingness to use a vaccination certificate in both use cases U1 and U2. This might be due to the fact that participants value the German Corona-Warn-App as an important instrument to help fight the pandemic and are more open for other measures in this regard. This also shows that participants might not see the need for other apps, as they are using the CWA already. 

When comparing the specific app-based vaccination certificate (C3) to the more generalized and feature-rich CWA-integrated vaccination certificate (C5) we do not find a significant difference, in fact in use case U3 there is no difference at all in the mean willingness to use values for those conditions. Since it is announced that the European vaccination certificate will be integrated into the CWA, we expect that many users already thought of this option and will use the existing CWA for this purpose. This also shows that it can make sense to add features to existing pandemic apps and develop them further in order to create additional incentives to continue using them.

\subsection{Design Implications for Vax Apps}
Our results show important factors for the acceptance of vaccination certificates. Based on that we pose the following design implications for Vaccination certificates. 

Like other technology acceptance literature~\cite{davis_informationtechnology_1989, venkatesh_acceptance_2003} we also found factors as perceived usefulness and ease of use to highly correlate with the willingness to use the different variants of vaccination certificates as mentioned in Section~\ref{sec:effortandutility}. 
Other related work identified privacy concerns to be barriers for the use and acceptance of mobile health apps \cite{wottrich_privacy_2018, zhou_barriers_2019}. Our results confirm these findings. We found a higher privacy disposition to negatively influence the willingness to use vaccination certificates - paper-based and app-based ones. This is also in line with the development of the popular German contact tracing app (CWA) where great importance was attached to anonymity, a decentralized approach and privacy. Especially for sensitive health data in terms of apps, privacy plays an important role and should therefore strictly follow privacy-by-design principles. This should also be taken into account when the vaccination status is verified by third-parties. If possible, third-parties should only be able to see the (electronically signed) QR-code to make the process least privacy-invasive for the user. Therefore, we recommend easy of use, utility and high privacy standards to be key factors for the design of vaccination apps.

In our opinion, it is much more difficult to pursue anonymity and a decentralized approach with vaccination certificates, as is partly the case with contact tracing apps. However, vaccination certificate apps should follow a least-invasive privacy approach.
As participants value a high utility of the vaccination certificates, represented in both our quantitative and qualitative results, integrated and universal solutions should be considered preferable. 
However, universal compatibility, \eg, across multiple countries, requires a certain level of verifiability.
For the planned European digital vaccination certificate this should be taken into account especially for purposes such as Europe-wide and international traveling and cross-border commuters.

The app does not necessarily has to be coronavirus-specific but can rather be integrated as an additional feature in existing privacy-friendly solutions, as is the case with the Corona-Warn-App. As a design implication it makes sense to add features to existing pandemic apps and develop them further in order to create additional incentives to continue using them. 

Another important factor when designing digital and paper-based vaccination certificates is forgery protection, especially in the case of a pandemic to protect their lives and health of the population. Although our study did not focus on security, we will highlight some implications briefly. Especially paper-based documents are easy to forge compared with digital apps. In particular, the currently used yellow vaccination certificate lacks of any security features, such as hologram stickers. Strong forgery-protection could be a huge advantage of digital vaccination certificates, if implemented properly. We therefore recommend to implement forgery-protection into the design process.

Since our study showed that paper-based certificate variants also have a high willingness to use, it should nevertheless be possible to use these variants as an alternative to digital apps. In addition to that, some participants stated that they do not own a smartphone or are unable to install such an app for technical reasons. Therefore, we recommend that the app variant should, especially when considering forgery protection, be the preferred variant but paper-based alternatives should always be offered.

%% file: sections/6-limitations.tex
\section{Limitations}
Although we tried to make our scenarios and the questions contained therein as understandable and realistic as possible including a comparatively large pilot study in order to get the best possible quantitative and qualitative data, our study has the following limitations.

First, in the survey we present fictitious vaccination certificates to the participants (albeit based strongly on real-world prototypes and/or deployed ones).
Due to the dynamic situation with rapid changes, the examined certificates do not necessarily correspond in all details to these existing or planned vaccination certificates.

Second, an online survey will never be able to fully capture the complexity of interacting in real-world situations, possibly underestimating the influence of factors such as forgetting or loosing a paper-based certificate or an empty battery for digital versions.  We tried to mitigate these effects by having a clear explanation of the certificate on the survey, and adding a visual representation as well.  In addition, some conditions such as the WHO vaccination passport (C2) and the Corona-Warn-App (C5) are in wide-spread use, so several participants should be quite familiar with them. 
Future work may complement our survey by conducting field studies evaluating and using specific vaccination certificates including an app-based verification process in real-world applications which allows to observe participants' actual behavior within such scenarios.

Finally, at the time conducting the survey vaccination certificates were only used to document vaccinations, the possibility of easing restrictions for vaccinated were only discussed.
Thus, it is not clear to what extent there is a discrepancy between self-reported and actual behavior if vaccination certificates are required in real-world applications. 

%% file: sections/7-conclusion.tex
\section{Conclusion}
Vaccinations are the most promising measure to end the ongoing pandemic in the long run and to reduce restrictions under the use of vaccination certificates in the short term. 
For these certificates to meet their purpose and to be widely adopted it is important to take people's opinions into account. 
We therefore studied different solutions of vaccination certificates and participants' willingness to use them for different use cases. Overall, our results indicate that both the general willingness to use and utility of vaccination certificates are perceived positively across all investigated scenarios. 
Overall, paper-based solutions received more positive responses compared to app-based solutions. 
Important factors influencing the adoption of vaccination certificates include privacy concerns, worries about a coronavirus infection and positive attitudes towards vaccination. Appropriate handling of sensitive health data, and addressing privacy and security concerns are key factors for design and development for certificate apps. We conclude that for documentation only there is no need for an app-based variant but for verifying the vaccination status we recommend to offer both paper- and app-based vaccination certificates in a privacy preserving manner.

%% file: sections/questionnaire.tex
\section{QUESTIONNAIRE}
\label{sec:questionnaire}
\renewcommand{\labelitemi}{$\bullet$}

\paragraph*{Smartphone Use and Experiences With the Coronavirus}

\begin{enumerate}
    \item[Q1:] Do you own a smartphone? [single choice]
\begin{itemize}
    \item Yes; No
\end{itemize}
\end{enumerate}

\begin{enumerate}
    \item[Q2:] Do you use an app (or smartwatch) to monitor your health or track your fitness? [single choice]
\begin{itemize}
    \item Yes; No; Don’t know; Prefer not to answer
\end{itemize}
\end{enumerate}
\begin{enumerate}
    \item[Q3:] Do you use the official German Corona-Warn-App on your smartphone? [single choice]
\begin{itemize}
    \item Yes; No; I had installed the app, but have since deleted it again; Don’t know; Prefer not to answer
\end{itemize}
\end{enumerate}
\begin{enumerate}
    \item[Q4:] Are you or have you been infected with the novel coronavirus? [single choice]
\begin{itemize}
    \item Yes; No; Don’t know; Prefer not to answer
\end{itemize}
\end{enumerate}
\begin{enumerate}
    \item[Q5:] Is there a person in your social circle who is or has been infected with the coronavirus? [single choice]
\begin{itemize}
    \item Yes; No; Prefer not to answer
\end{itemize}
\end{enumerate}
\begin{enumerate}
    \item[Q6:] How concerned are you that you will become infected with the coronavirus? [single choice]
\begin{itemize}
    \item 1 -- Not concerned; 2 -- A-little concerned; 3 -- Moderately concerned; 4 -- Quite-a-bit concerned; 5 -- Very concerned
    \item Prefer not to answer
\end{itemize}
\end{enumerate}
\begin{enumerate}
    \item[Q7:] How concerned are you that someone you are close to may be infected with the coronavirus? [single choice]; same answer options as Q6
\end{enumerate}
\paragraph*{Vaccinations}
\begin{enumerate}
    \item[Q8:] Do you possess the pictured yellow vaccination card? [single choice]
\begin{itemize}
    \item Yes; No; Don’t know; Prefer not to answer
\end{itemize}
\end{enumerate}
\begin{enumerate}
    \item[Q9:] Do you know where your vaccination card is right now? [single choice]
\begin{itemize}
    \item Yes; No; Don’t know; Prefer not to answer
\end{itemize}
\end{enumerate}
\begin{enumerate}
    \item[Q10:] Do you have the vaccinations recommended by the STIKO (Standing Committee on Vaccination), for example, against measles, tetanus, and influenza (flu)? [single choice]
\begin{itemize}
    \item Yes; No; Don’t know; Prefer not to answer
\end{itemize}
\end{enumerate}
\begin{enumerate}
    \item[Q11:] What is the reason for that? [multiple choice]
\begin{itemize}
    \item I forget my vaccination appointments (occasionally); I did not read up on the recommended vaccinations; I generally do not want to be vaccinated; I am afraid of possible side effects; I want my body to fight the virus infection itself; Don’t know; Prefer not to answer
\end{itemize}
\end{enumerate}
\begin{enumerate}
    \item[Q12:] Have you already been vaccinated against the coronavirus? [single choice]
\begin{itemize}
    \item Yes; No; Prefer not to answer
\end{itemize}
\end{enumerate}
\begin{enumerate}
    \item[Q13:] Would you like to be vaccinated against the coronavirus? [if Q12=="No"; single choice]
\begin{itemize}
    \item Yes; No; Prefer not to answer
\end{itemize}
\end{enumerate}
\begin{enumerate}
    \item[Q14:] Please indicate your agreement for the following statement: Most people I care about think I should get vaccinated against the coronavirus. [single choice]
\begin{itemize}
    \item 1 -- Fully-disagree; 2 -- Mainly-disagree; 3 -- Neutral; 4 -- Mainly-agree; 5 -- Fully-agree
\end{itemize}
\end{enumerate}
\begin{enumerate}
    \item[Q15:] Is there a person in your personal circle who has already been vaccinated against the coronavirus? [single choice]
\begin{itemize}
    \item Yes; No; Don’t know; Prefer not to answer
\end{itemize}
\end{enumerate}
\paragraph*{Current Situation Description}
To ensure that all study participants are on the same level of knowledge for the following questions, we have provided some brief information for you. Please read them and keep in mind the current situation for answering them: In Germany, people have been vaccinated against coronavirus since December 27, 2020. For a full vaccination protection, two vaccination appointments at a time interval are necessary. Since there is currently only limited vaccine available, vaccination is carried out according to a staged plan. After the completed vaccination, a confirmation of vaccination is issued.
On the next pages you will find an example of such a confirmation. The following questions refer to the described scenario.
\begin{enumerate}
    \item[Q17:] Have you read the situation description carefully? [single choice]; Yes
\end{enumerate}
\paragraph*{[Variant of Vaccination Certificate; use case U1]}
[Each of the following question groups vary minimally due to the five studied conditions.]
Imagine a personalized paper-based vaccination confirmation is being issued after you have received your COVID-19 vaccination.
An example of a possible vaccination confirmation is shown in the following figure and includes the following information:
\begin{enumerate}
  \item Name and date of birth of the vaccinated person.
  \item Location of the vaccination center.
  \item Vaccine
  \item Day as well as confirmation of both performed necessary vaccinations.
  \item QR-code to verify the performed vaccinations. [except for WHO Certificate]
\end{enumerate}
\begin{enumerate}
    \item[Q18:] Would you have this confirmation issued? [single choice]
\begin{itemize}
    \item 1 -- Certainly-not; 2 -- Unlikely; 3 -- About-50:50; 4 -- Likely; 5 -- For-sure
\end{itemize}
\end{enumerate}
\begin{enumerate}
    \item[Q19:] Why would you (probably) not have this confirmation issued? [if Q18==1 or Q18==2; free text]
\end{enumerate}
\begin{enumerate}
    \item[Q20:] Why are you undecided about whether you would have this confirmation issued? [if Q18==3; free text]
\end{enumerate}
\begin{enumerate}
    \item[Q21:] Why would you (probably) have this confirmation issued? [if Q18==4 or Q18==5; free text]
\end{enumerate}
\paragraph*{Proof of Vaccination}
Imagine that the issued [condition-based] vaccination certificate can also be used to \textbf{proof your vaccination status}. This should allow additional relaxation, prevent future lock-downs, and curb the spread of the coronavirus.
People with relevant pre-existing conditions, pregnant women, and children are exempt from this required proof of vaccination against the coronavirus, i.\,e., generally people for whom vaccination is not recommended from a medical perspective. Please keep in mind the current pandemic and vaccination situation as described previously.
\begin{enumerate}
    \item[Q22:] Have you read the description text carefully? [single choice]; Yes
\end{enumerate}

\paragraph*{[Variant of Vaccination Certificate; use case U2]}
\begin{enumerate}
    \item[Q23:] Imagine that proof-of-vaccination against the coronavirus becomes necessary for various purposes -- both in Germany and worldwide. For each of the following statements, please indicate the extent to which you agree with it: The [condition] vaccination certificate should be shown to\ldots [array of single choice questions; answer option for each: 1 -- Fully-disagree, 2 -- Mainly-disagree, 3 -- Neutral, 4 -- Mainly-agree, 5 -- Fully-agree]
\begin{itemize}
    \item \ldots to be allowed to travel \underline{nationally} via airplane. ; \underline{internationally} via airplane. ; \underline{nationally} via train. ; \underline{internationally} via train.
    \item \ldots to be allowed to enter other countries by car.
    \item \ldots to stay overnight in hotels (national and abroad).
    \item \ldots to be allowed to participate in major events (such as soccer matches and concerts).
    \item \ldots to visit places such as restaurants, museums, and cinemas.
    \item \ldots to visit schools and kindergartens or community facilities such as daycare and after-school programs.
    \item \ldots to work in professions with public traffic such as hospitals, offices, and supermarkets.
    \item \ldots to be allowed to work in offices which are simultaneously used by several people and less than 10 square meters per person. [ruled by the occupational health and safety regulation in Germany]
\end{itemize}
\end{enumerate}
\begin{enumerate}
    \item[Q24:] Would you use the vaccination certificate to proof your vaccination status, for example, for some of the purposes mentioned previously? [single choice]
\begin{itemize}
    \item 1 -- Certainly-not; 2 -- Unlikely; 3 -- About-50:50; 4 -- Likely; 5 -- For-sure
\end{itemize}
\end{enumerate}
\begin{enumerate}
    \item[Q25:] Please indicate your agreement for the following statement: The use of this vaccination certificate as proof-of-vaccination is useful to allow early relaxation of current measures and to prevent future lock-downs. [single choice]
\begin{itemize}
    \item 1 -- Fully-disagree; 2 -- Mainly-disagree; 3 -- Neutral; 4 -- Mainly-agree; 5 -- Fully-agree
\end{itemize}
\end{enumerate}
\begin{enumerate}
    \item[Q26:] How effortful do you consider the use of the vaccination certificate to verify your vaccination status, for example, for some of the mentioned purposes? [single choice]
\begin{itemize}
    \item 1 -- Not effortful; 2 -- A-little effortful; 3 -- Moderately effortful; 4 -- Quite-a-bit effortful; 5 -- Very effortful
\end{itemize}
\end{enumerate}
\paragraph*{Proof of Vaccination With Fictitious Scenario}
\textbf{Attention!} The following questions now refer to a \textbf{fictitious} scenario [use case U3] described below. Please read this scenario [use case U3] carefully. Please empathize the scenario as best you can and answer the questions on this basis. The situation in Germany is now as follows:
\textbf{Every German citizen has received an offer of vaccination, there was and is enough vaccine in stock}. However, the pandemic is not over yet. Since vaccination is not compulsory and vaccinated people are still potentially contagious, various measures to combat the pandemic are still in effect. 
\begin{enumerate}
    \item[Q27:] Based on the description above, please select the correct statement. [single choice; attention-check]
\begin{itemize}
    \item It is assumed that there is enough vaccine for all citizens; Limited vaccine is believed to be available.
\end{itemize}
\end{enumerate}
\paragraph*{[Variant of Vaccination Certificate; use case U3]}
\begin{enumerate}
    \item[Q28:] Same question as Q23 [in use case U2].
\end{enumerate}
\begin{enumerate}
    \item[Q29:] Same question as Q24 [in use case U2].
\end{enumerate}
\begin{enumerate}
    \item[Q30:] What is the reason for that? [free text]
\end{enumerate}
\begin{enumerate}
    \item[Q31:] Same question as Q25 [in use case U2].
\end{enumerate}
\begin{enumerate}
    \item[Q32:] Would you generally use a vaccination certificate? If yes, in what form? [single choice]
\begin{itemize}
    \item Yes, a digital proof via app.
    \item Yes, a paper-based proof.
    \item Don’t know
    \item No, I would not use a vaccination certificate at all.
\end{itemize}
\end{enumerate}
\begin{enumerate}
    \item[Q33:] Would you support a mandatory vaccination against the coronavirus in Germany? [single choice]
\begin{itemize}
    \item Yes; No; Don’t know; Prefer not to answer
\end{itemize}
\end{enumerate}
\begin{enumerate}
    \item[Q34:] Please indicate your agreement with the following statement: Personally, I feel strongly burdened by the coronavirus pandemic. [single choice]
\begin{itemize}
    \item 1 -- Fully-disagree; 2 -- Mainly-disagree; 3 -- Neutral; 4 -- Mainly-agree; 5 -- Fully-agree
\end{itemize}
\end{enumerate}
\paragraph*{Privacy Disposition}
\begin{enumerate}
    \item[Q35:] For each of the following statements, please indicate the extent to which you agree.\footnote{The first three items are from the ``Disposition to privacy'' scale in the version of Yuan Li \cite{Li_2014_privacy}.} [array of single choice questions; answer option for each: 1 -- Fully-disagree, 2 -- Mainly-disagree, 3 -- Neutral, 4 -- Mainly-agree, 5 -- Fully-agree]
\begin{itemize}
    \item Compared to others, I am more sensitive about the way other people or organizations handle my personal information. 
    \item Compared to others, I see more importance in keeping personal information private. 
    \item Compared to others, I am less concerned about potential threats to my personal privacy. (R) 
    \item Compared to others, I value health data as especially worthy of protection. 
\end{itemize}
\end{enumerate}
\begin{enumerate}
    \item[Q36:] For each of the following statements, please indicate the extent to which you agree.\footnote{The first three items are from the``Perceived Privacy Risk'' scale in the version of Chen and Cai \cite{Chen_2012_PrivacyApps}.} [array of single choice questions; answer option for each: 1 -- Fully-disagree, 2 -- Mainly-disagree, 3 -- Neutral, 4 -- Mainly-agree, 5 -- Fully-agree; question is shown in App-based conditions \textbf{only}]
\begin{itemize}
    \item I am concerned that the information I submit in this app could be misused.
    \item I am concerned about submitting information in this app, because of what others might do with it
    \item I am concerned about submitting information in this app, because it could be be used in a way I did not foresee.
    \item I am concerned about disclosing health data in this app.
\end{itemize}
\end{enumerate}
\paragraph*{Demographics}
\begin{enumerate}
    \item[Q37--Q40:] Gender, Age, Education, previous IT-knowledge
\end{enumerate}

%% file: proof-of-vax.bbl
% Generated by IEEEtranS.bst, version: 1.14 (2015/08/26)
\begin{thebibliography}{10}
\providecommand{\url}[1]{#1}
\csname url@samestyle\endcsname
\providecommand{\newblock}{\relax}
\providecommand{\bibinfo}[2]{#2}
\providecommand{\BIBentrySTDinterwordspacing}{\spaceskip=0pt\relax}
\providecommand{\BIBentryALTinterwordstretchfactor}{4}
\providecommand{\BIBentryALTinterwordspacing}{\spaceskip=\fontdimen2\font plus
\BIBentryALTinterwordstretchfactor\fontdimen3\font minus
  \fontdimen4\font\relax}
\providecommand{\BIBforeignlanguage}[2]{{%
\expandafter\ifx\csname l@#1\endcsname\relax
\typeout{** WARNING: IEEEtranS.bst: No hyphenation pattern has been}%
\typeout{** loaded for the language `#1'. Using the pattern for}%
\typeout{** the default language instead.}%
\else
\language=\csname l@#1\endcsname
\fi
#2}}
\providecommand{\BIBdecl}{\relax}
\BIBdecl

\bibitem{altmann_acceptability_2020}
S.~Altmann, L.~Milsom, H.~Zillessen, R.~Blasone, F.~Gerdon, R.~Bach,
  F.~Kreuter, D.~Nosenzo, S.~Toussaert, and J.~Abeler, ``{Acceptability of
  App-Based Contact Tracing for COVID-19: Cross-Country Survey Study},''
  \emph{JMIR mHealth and uHealth}, vol.~8, no.~8, p. e19857, Aug. 2020.

\bibitem{baig_dnatesting_2020}
K.~Baig, R.~Mohamed, A.-L. Theus, and S.~Chiasson, ``{``I'm hoping they're an
  ethical company that won't do anything that I'll regret'': Users Perceptions
  of At-home DNA Testing Companies},'' in \emph{Proceedings of the 2020 CHI
  Conference on Human Factors in Computing Systems}, ser. CHI 2020.\hskip 1em
  plus 0.5em minus 0.4em\relax New York, NY, USA: ACM, 2020, pp. 1--13.

\bibitem{bol19:differences-mobile-health}
N.~Bol, N.~Helberger, and J.~C.~M. Weert, ``{Differences in Mobile Health App
  Use: A Source of new Digital Inequalities?}'' \emph{The Information Society},
  vol.~34, no.~3, pp. 183--193, Apr. 2018.

\bibitem{cccsocialimplications_2021}
{Chaos Computer Club}, ``{Impfausweise beenden keine Pandemien},'' May 2021,
  \url{https://www.ccc.de/en/updates/2021/impfausweise-beenden-keine-pandemien}
  as of \today.

\bibitem{germanynotenoughvaccine_2021}
B.~Chappell, ``{Not Enough Vaccine Doses In Europe To Stop A 3rd Wave, German
  Health Minister Says},'' Mar. 2021,
  \url{https://www.npr.org/sections/coronavirus-live-updates/2021/03/19/979146521/not-enough-vaccine-doses-in-europe-to-stop-a-3rd-wave-german-health-minister-say}
  as of \today.

\bibitem{Chen_2012_PrivacyApps}
X.~Chen and S.~Cai, ``Self-disclosure under social networking sites: A
  risk-utility decision model,'' in \emph{International Conference on
  Electronic Commerce}.\hskip 1em plus 0.5em minus 0.4em\relax Association for
  Computing Machinery, 2012.

\bibitem{davidson_healthcode_2020}
H.~Davidson, ``{China's coronavirus health code apps raise concerns over
  privacy},'' Apr. 2020,
  \url{https://www.theguardian.com/world/2020/apr/01/chinas-coronavirus-health-code-apps-raise-concerns-over-privacy}
  as of \today.

\bibitem{davis_informationtechnology_1989}
\BIBentryALTinterwordspacing
F.~D. Davis, ``{Perceived Usefulness, Perceived Ease of Use, and User
  Acceptance of Information Technology},'' \emph{MIS Quarterly}, vol.~13,
  no.~3, pp. 319--340, Sep. 1989. [Online]. Available:
  \url{https://www.jstor.org/stable/249008}
\BIBentrySTDinterwordspacing

\bibitem{germanysecondlockdownpoorly_2020}
L.~Eberle, M.~Feldenkirchen, M.~Hassenkamp, M.~Knobbe, V.~Medick, M.~Rosenbach,
  L.~Rosenfelder, and G.~Traufetter, ``{Germany Is Faring Poorly in the Second
  Wave of the Coronavirus},'' Dec. 2020,
  \url{https://www.spiegel.de/international/germany/germany-is-faring-poorly-in-the-second-wave-of-the-coronavirus-a-afc634db-9220-496b-8aca-fd19f2e0962f}
  as of \today.

\bibitem{vaccinepassport_2021}
C.~Elliott, ``What you need to know about vaccine passports,'' Dec. 2020,
  \url{https://www.washingtonpost.com/lifestyle/travel/yellow-card-vaccine-passport/2020/12/30/746c0558-40b7-11eb-8db8-395dedaaa036_story.html}
  as of \today.

\bibitem{emami-naeini21:which-privacy-security}
P.~Emami-Naeini, J.~Dheenadhayalan, Y.~Agarwal, and L.~F. Cranor, ``{Which
  Privacy and Security Attributes Most Impact Consumers' Perception and
  Willingness to Purchase IoT Devices?}'' in \emph{IEEE Symposium on Security
  and Privacy}, ser. SP~'21.\hskip 1em plus 0.5em minus 0.4em\relax San
  Francisco, CA, USA: IEEE, May 2021.

\bibitem{eu21:digital-green-certificate}
{European Commission}, ``{Coronavirus: Commission proposes a Digital Green
  Certificate},'' May 2021,
  \url{https://ec.europa.eu/commission/presscorner/detail/en/IP_21_1181/} as of
  \today.

\bibitem{europeandigitalvaxcertificate_2021}
------, ``Covid-19: Digital green certificates,'' May 2021,
  \url{https://ec.europa.eu/info/live-work-travel-eu/coronavirus-response/safe-covid-19-vaccines-europeans/covid-19-digital-green-certificates_en}
  as of \today.

\bibitem{gerlatestcovidupdates_2020}
{FAZIT Communication GmbH and Federal Foreign Office}, ``Latest coronavirus
  updates,'' May 2021,
  \url{https://www.deutschland.de/en/news/coronavirus-in-germany-informations}
  as of \today.

\bibitem{federalministryepa_2020}
{German Federal Ministry of Health}, ``{Driving the digital transformation of
  Germany’s healthcare system for the good of patients},'' Dec. 2020,
  \url{https://www.bundesgesundheitsministerium.de/en/en/topics/digital-healthcare-act-dvg.html}
  as of \today.

\bibitem{personalelectronichealth_2021}
M.~Grote-Westrick, ``{New German digital project paves the way for online
  access to personal electronic health records},'' Feb. 2021,
  \url{https://blogs.bmj.com/bmj/2021/02/18/new-german-digital-project-paves-the-way-for-online-access-to-personal-electronic-health-records/}
  as of \today.

\bibitem{gu_privacy_2017}
J.~Gu, Y.~C. Xu, H.~Xu, C.~Zhang, and H.~Ling, ``{Privacy concerns for mobile
  app download: An elaboration likelihood model perspective},'' \emph{Decision
  Support Systems}, vol.~94, pp. 19--28, Feb. 2017.

\bibitem{harborth19:explaining-technology-use}
D.~Harborth, S.~Pape, and K.~Rannenberg, ``{Explaining the Technology Use
  Behavior of Privacy-Enhancing Technologies: The Case of Tor and JonDonym},''
  in \emph{Proceedings on Privacy Enhancing Technologies 2020}.\hskip 1em plus
  0.5em minus 0.4em\relax Sciendo, Dec. 2019, pp. 111--128.

\bibitem{horowitz_europe_2020}
J.~Horowitz and A.~Satariano, ``{Europe Rolls Out Contact Tracing Apps, With
  Hope and Trepidation},'' Jun. 2020,
  \url{https://www.nytimes.com/2020/06/16/world/europe/contact-tracing-apps-europe-coronavirus.html}
  as of \today.

\bibitem{rki_downloadzahlen_2020}
R.~K. Institut, ``{Kennzahlen zur CORONA-WARN-APP},'' May 2021,
  \url{https://www.rki.de/DE/Content/InfAZ/N/Neuartiges_Coronavirus/WarnApp/Archiv_Kennzahlen/Kennzahlen_28052021.pdf?__blob=publicationFile}
  as of \today.

\bibitem{iata21:digital-travel-pass}
{International Air Transport Association}, ``Iata travel pass initiative,'' May
  2021, \url{https://www.iata.org/en/programs/passenger/travel-pass/} as of
  \today.

\bibitem{europe21:importanceofdigitalhealthdata}
------, ``{The Rise of Digital Health Technologies During the Pandemic},'' Apr.
  2021,
  \url{https://www.europarl.europa.eu/RegData/etudes/BRIE/2021/690548/EPRS_BRI(2021)690548_EN.pdf}
  as of \today.

\bibitem{ICCESOMAR_Code_2021}
\BIBentryALTinterwordspacing
{International Chamber of Commerce} and {European Society for Opinion and
  Marketing Research}, ``{International Code on Market and Social Research},''
  {ICC/ESOMAR}, Tech. Rep., Dec. 2007. [Online]. Available:
  \url{https://iccwbo.org/publication/iccesomar-international-code-on-market-and-social-research/}
\BIBentrySTDinterwordspacing

\bibitem{israel21:digital-green-pass}
{Israel Ministry of Health}, ``{What is a Green Pass?}'' May 2021,
  \url{https://corona.health.gov.il/en/directives/green-pass-info/} as of
  \today.

\bibitem{germandebateprivileges_2021}
F.~Jordans, ``Germany debates privileges for those who've been vaccinated,''
  Apr. 2021,
  \url{https://abcnews.go.com/Health/wireStory/germany-debates-privileges-whove-vaccinated-77318926}
  as of \today.

\bibitem{kaptchuk_covid19apps_2020}
G.~Kaptchuk, D.~G. Goldstein, E.~Hargittai, J.~Hofman, and E.~M. Redmiles,
  ``{How Good is Good Enough for COVID19 Apps? The Influence of Benefits,
  Accuracy, and Privacy on Willingness to Adopt},'' \emph{arXiv preprint
  arXiv:2005.04343}, 2020.

\bibitem{li_covid_apps_2020}
T.~Li, J.~J. Yang, C.~Faklakis, J.~King, Y.~Agarwal, L.~Dabbish, and J.~I.
  Hong, ``{Decentralized is not risk-free: Understanding public perceptions of
  privacy-utility trade-offs in COVID-19 contact-tracing apps},'' \emph{arXiv
  preprint arXiv:2005.11957}, 2020.

\bibitem{Li_2014_privacy}
Y.~Li, ``A multi-level model of individual information privacy beliefs,''
  \emph{Electronic Commerce Research and Application}, vol.~13, no.~1, 2014.

\bibitem{linsner21:role-privacy-digitalization}
S.~Linsner, F.~Kuntke, E.~Steinbrink, J.~Franken, and C.~Reuter, ``{The Role of
  Privacy in Digitalizaton -- Analyzing Perspectives of German Farmers},'' in
  \emph{Proceedings on Privacy Enhancing Technologies 2021}.\hskip 1em plus
  0.5em minus 0.4em\relax Sciendo, Mar. 2021, pp. 334--350.

\bibitem{lu21:comparing-perspectives}
X.~Lu, T.~L. Reynolds, E.~Jo, H.~Hong, X.~Page, Y.~Chen, and D.~A. Epstein,
  ``{Comparing Perspectives Around Human and Technology Support for Contact
  Tracing},'' in \emph{ACM CHI Conference on Human Factors in Computing
  Systems}, ser. CHI~'21.\hskip 1em plus 0.5em minus 0.4em\relax Virtual Event:
  ACM, May 2021.

\bibitem{mathieu21:global-database-covid}
E.~Mathieu, H.~Ritchie, E.~Ortiz-Ospina, M.~Roser, J.~Hasell, C.~Appel,
  C.~Giattino, and L.~Rod\'{e}s-Guirao, ``{A Global Database of COVID-19
  Vaccinations},'' \emph{Nature Human Behaviour}, 2021.

\bibitem{mathieu21:vax-data}
------, ``{Coronavirus (COVID-19) Vaccinations},'' May 2021,
  \url{https://ourworldindata.org/covid-vaccination} as of \today.

\bibitem{cdc_ukproofofvax_2020}
B.~Mueller and C.~Zimmer, ``U.k. coronavirus vaccine: Side effects, safety, and
  who gets it first,'' Dec. 2020,
  \url{https://www.nytimes.com/2020/12/08/world/europe/coronavirus-vaccine-uk-questions.html}
  as of \today.

\bibitem{namara19:emotional-practical-considerations}
M.~Namara, D.~Wilkinson, K.~Caine, and B.~P. Knijnenburg, ``{Emotional and
  Practical Considerations Towards the Adoption and Abandonment of VPNs as
  Privacy-Enhancing Technology},'' in \emph{Proceedings on Privacy Enhancing
  Technologies 2020}.\hskip 1em plus 0.5em minus 0.4em\relax Sciendo, Sep.
  2019, pp. 83--102.

\bibitem{nicholas_mentalhealth_2019}
J.~Nicholas, K.~Shilton, S.~M. Schueller, E.~L. Gray, M.~J. Kwasny, and D.~C.
  Mohr, ``{The Role of Data Type and Recipient in Individuals’ Perspectives
  on Sharing Passively Collected Smartphone Data for Mental Health:
  Cross-Sectional Questionnaire Study},'' \emph{JMIR mHealth and uHealth},
  vol.~7, no.~4, p. e12578, Apr. 2019.

\bibitem{belmontreport_2018}
\BIBentryALTinterwordspacing
{Office for Human Research Protections}, ``{The Belmont Report},'' Office for
  Human Research Protections (OHRP), Tech. Rep., Jan. 2018. [Online].
  Available:
  \url{https://www.hhs.gov/ohrp/regulations-and-policy/belmont-report/read-the-belmont-report/index.html}
\BIBentrySTDinterwordspacing

\bibitem{germandebateprivilegesunfair_2021}
P.~Oltermann, ``German 'jab to freedom’ covid bill criticised as unfair to
  young people,'' May 2021,
  \url{https://www.theguardian.com/world/2021/may/04/german-jab-to-freedom-covid-bill-criticised-as-unfair-to-young-people}
  as of \today.

\bibitem{peng_healthapps_2016}
W.~Peng, S.~Kanthawala, S.~Yuan, and S.~A. Hussain, ``{A qualitative study of
  user perceptions of mobile health apps},'' \emph{BMC Public Health}, vol.~16,
  no. 1158, pp. 1--11, Nov. 2016.

\bibitem{rasche18:prevalence-health-app}
P.~Rasche, M.~Wille, C.~Br\"{o}hl, S.~Theis, K.~Sch\"{a}fer, M.~Knobe, and
  A.~Mertens, ``{Prevalence of Health App Use Among Older Adults in Germany:
  National Survey},'' \emph{JMIR mHealth and uHealth}, vol.~6, no.~1, Jan.
  2018.

\bibitem{ray20:warn-them-just}
H.~Ray, F.~Wolf, R.~Kuber, and A.~J. Aviv, ``{``Warn Them'' or ``Just Block
  Them''?: Investigating Privacy Concerns Among Older and Working Age
  Adults},'' in \emph{Proceedings on Privacy Enhancing Technologies
  2021}.\hskip 1em plus 0.5em minus 0.4em\relax Sciendo, Dec. 2020, pp. 27--47.

\bibitem{rki_covid19dashboard_2020}
{Robert Koch Institut}, ``{COVID-19 Dashboard},'' May 2021,
  \url{https://experience.arcgis.com/experience/478220a4c454480e823b17327b2bf1d4}
  as of \today.

\bibitem{germanyvaccinemonitoring_2021}
------, ``{Digitales Impfquotenmonitoring zur COVID-19-Impfung},'' May 2021,
  \url{https://www.rki.de/DE/Content/InfAZ/N/Neuartiges_Coronavirus/Daten/Impfquoten-Tab.html}
  as of \today.

\bibitem{covpass_2021}
------, ``{The CovPass-App},'' May 2021,
  \url{https://digitaler-impfnachweis-app.de/} as of \today.

\bibitem{rohrmann78:empirische-studien-entwicklung}
B.~Rohrmann, ``{Empirische Studien zur Entwicklung von Antwortskalen f\"{u}r
  die sozialwissenschaftliche Forschung.}'' \emph{Zeitschrift f\"{u}r
  Sozialpsychologie}, vol.~9, pp. 222--245, 1978.

\bibitem{rohrmann07:verbal-qualifiers-rating}
------, ``{Verbal Qualifiers for Rating Scales: Sociolinguistic Considerations
  and Psychometric Data},'' University of Melbourne, Melbourne, Australia,
  Tech. Rep., 2007.

\bibitem{sap_coronawarnapp_2020}
{SAP Germany}, ``{Open-Source Project Corona-Warn-App},'' {SAP Germany}, Jun.
  2020, \url{https://www.coronawarn.app/en/} as of \today.

\bibitem{schnitzler20:managing-longitudinal-privacy}
T.~Schnitzler, S.~Mirza, M.~D{\"u}rmuth, and C.~P{\"o}pper, ``{SoK: Managing
  Longitudinal Privacy of Publicly Shared Personal Online Data},'' in
  \emph{Proceedings on Privacy Enhancing Technologies 2021}.\hskip 1em plus
  0.5em minus 0.4em\relax Sciendo, Nov. 2020, pp. 229--249.

\bibitem{simko_contact_tracing_2020}
L.~Simko, R.~Calo, F.~Roesner, and T.~Kohno, ``{COVID-19 Contact Tracing and
  Privacy: Studying Opinion and Preferences},'' \emph{arXiv preprint
  arXiv:2005.06056}, 2020.

\bibitem{story21:awareness-adoption-misconceptions}
P.~Story, D.~Smullen, Y.~Yao, A.~Acquisti, L.~F. Cranor, N.~Sadeh, and
  F.~Schaub, ``{Awareness, Adoption, and Misconceptions of Web Privacy
  Tools},'' in \emph{Proceedings on Privacy Enhancing Technologies 2021}.\hskip
  1em plus 0.5em minus 0.4em\relax Sciendo, Mar. 2021, pp. 308--333.

\bibitem{trang_oneapp_2020}
\BIBentryALTinterwordspacing
S.~Trang, M.~Trenz, W.~H. Weiger, M.~Tarafdar, and C.~M.~K. Cheung, ``{One app
  to trace them all? Examining app specifications for mass acceptance of
  contact-tracing apps},'' \emph{European Journal of Information Systems},
  vol.~29, no.~3, pp. 1--14, Jul. 2020. [Online]. Available:
  \url{https://doi.org/10.1080/0960085X.2020.1784046}
\BIBentrySTDinterwordspacing

\bibitem{tsai21:exploring-promoting-diagnostic}
C.-H. Tsai, Y.~You, X.~Gui, Y.~Kou, and J.~M. Carroll, ``{Exploring and
  Promoting Diagnostic Transparency and Explainability in Online Symptom
  Checkers},'' in \emph{ACM CHI Conference on Human Factors in Computing
  Systems}, ser. CHI~'21.\hskip 1em plus 0.5em minus 0.4em\relax Virtual Event:
  ACM, May 2021.

\bibitem{us21:fully-vaccinated-privileges}
{US Centers of Disease Control and Prevention}, ``{When You've Been Fully
  Vaccinated},'' May 2021,
  \url{https://www.cdc.gov/coronavirus/2019-ncov/vaccines/fully-vaccinated.html}
  as of \today.

\bibitem{us21:vaccination-record-us}
{US Department of Defense}, ``{Vaccine Record Card},'' May 2021,
  \url{https://www.defense.gov/observe/photo-gallery/igphoto/2002536141/} as of
  \today.

\bibitem{utz21:apps-against-spread}
C.~Utz, S.~Becker, T.~Schnitzler, F.~Farke, F.~Herbert, L.~Schaewitz,
  M.~Degeling, and M.~D{\"u}rmuth, ``{Apps Against the Spread: Privacy
  Implications and User Acceptance of COVID-19-Related Smartphone Apps on Three
  Continents},'' in \emph{ACM CHI Conference on Human Factors in Computing
  Systems}, ser. CHI~'21.\hskip 1em plus 0.5em minus 0.4em\relax Virtual Event:
  ACM, May 2021.

\bibitem{venkatesh_acceptance_2003}
V.~Venkatesh, M.~G. Morris, G.~B. Davis, and F.~D. Davis, ``{User Acceptance of
  Information Technology: Toward a Unified View},'' \emph{MIS Quarterly},
  vol.~27, no.~3, pp. 425--478, Sep. 2003.

\bibitem{rasche21:green-pass-privileges}
R.~Wilf-Miron, V.~Myers, and M.~Saban, ``{Incentivizing Vaccination Uptake: The
  “Green Pass” Proposal in Israel},'' \emph{JAMA}, vol. 325, no.~15, Apr.
  2021.

\bibitem{heisesechealthapps_2020}
D.~Wischnjak, ``{IT-Sicherheit in der Medizin: 22 Krankenkassen-Apps im
  Sicherheits-Check},'' Dec. 2020,
  \url{https://www.heise.de/hintergrund/IT-Sicherheit-in-der-Medizin-22-Krankenkassen-Apps-im-Sicherheits-Check-4992896.html}
  as of \today.

\bibitem{internationalcertificatesofvax_2021}
{World Health Organization (WHO)}, ``Smart vaccination certificate working
  group,'' Dec. 2020,
  \url{https://www.who.int/groups/smart-vaccination-certificate-working-group}
  as of \today.

\bibitem{wottrich_privacy_2018}
V.~M. Wottrich, E.~A. {van Reijmersdal}, and E.~G. Smit, ``{The privacy
  trade-off for mobile app downloads: The roles of app value, intrusiveness,
  and privacy concerns},'' \emph{Decision Support Systems}, vol. 106, pp.
  44--52, Feb. 2018.

\bibitem{zhang_covid_privacy_2020}
B.~Zhang, S.~Kreps, and N.~McMurry, ``{Americans' Perceptions of Privacy and
  Surveillance in the COVID-19 Pandemic},'' \emph{OSF preprint osf.io/9wz3y},
  2020.

\bibitem{zhang21:mapping-landscape}
Y.~Zhang, Y.~Sun, L.~Padilla, S.~Barua, E.~Bertini, and A.~G. Parker,
  ``{Mapping the Landspace of COIVD-19 Crisis Visualizations},'' in \emph{ACM
  CHI Conference on Human Factors in Computing Systems}, ser. CHI~'21.\hskip
  1em plus 0.5em minus 0.4em\relax Virtual Event: ACM, May 2021.

\bibitem{zhou_barriers_2019}
L.~Zhou, J.~Bao, V.~Watzlaf, and B.~Parmanto, ``{Barriers to and Facilitators
  of the Use of Mobile Health Apps From a Security Perspective: Mixed-Methods
  Study},'' \emph{JMIR mHealth and uHealth}, vol.~7, no.~4, p. e11223, Apr.
  2019.

\end{thebibliography}
